\documentclass[journal,draftcls,onecolumn,12pt,twoside]{IEEEtran}
\usepackage[T1]{fontenc}
\usepackage[latin9]{inputenc}
\usepackage{amsthm}
\usepackage{amsmath} 
\usepackage{graphicx} 
\usepackage{color} 
\usepackage{cite}
\usepackage{algpseudocode}
\usepackage{algorithm}
\usepackage{amssymb}
\usepackage{subfigure}
\usepackage{esint}
\usepackage{algpseudocode}
\usepackage{epstopdf}
\usepackage{multirow}
\usepackage{makecell}
\usepackage{float}
\usepackage{lipsum}
\usepackage{balance}
\usepackage{ulem}
\usepackage{bbm}

\newtheorem{definition}{Definition}
\newtheorem{remark}{Remark}

\newtheorem{example}{Example}

\theoremstyle{plain}
\theoremstyle{plain}
\newtheorem{theorem}{Theorem}

\newcommand{\comment}[1]{}

\IEEEoverridecommandlockouts

\begin{document}

\title{Efficient Constrained Codes That Enable Page Separation in Modern Flash Memories}

\author{
   \IEEEauthorblockN{Ahmed Hareedy, \IEEEmembership{Member, IEEE}, Simeng Zheng, \IEEEmembership{Student Member, IEEE},\\ Paul Siegel, \IEEEmembership{Life Fellow, IEEE}, and Robert Calderbank, \IEEEmembership{Life Fellow, IEEE}}
   \thanks{This work was supported in part by the T\"{U}B\.{I}TAK 2232-B International Fellowship for Early Stage Researchers. This article was presented in part at the 2022 IEEE International Conference on Communications (ICC) \cite{ahh_rrconst}.
   
   Ahmed Hareedy is with the Department of Electrical and Electronics Engineering,
Middle East Technical University (METU), 06800 Ankara, Turkey (e-mail: ahareedy@metu.edu.tr).

   Simeng Zheng and Paul Siegel are with the Department of Electrical and Computer Engineering, University of California, San Diego (UCSD), La Jolla, CA 92093 USA (e-mail: sizheng@ucsd.edu; psiegel@ucsd.edu).
               
   Robert Calderbank is with the Department of Electrical and Computer Engineering, Duke University, Durham, NC 27708 USA (e-mail: robert.calderbank@duke.edu).}
}
\maketitle

\vspace{-2.7em}
\begin{abstract}

The pivotal storage density win achieved by solid-state devices over magnetic devices in 2015 is a result of multiple innovations in physics, architecture, and signal processing. One of the most important innovations in that regard is enabling the storage of more than one bit per cell in the Flash device, i.e., having more than two charge levels per cell. Constrained coding is used in Flash devices to increase reliability via mitigating inter-cell interference that stems from charge propagation among cells. Recently, capacity-achieving constrained codes were introduced to serve that purpose in modern Flash devices, which have more than two levels per cell. While these codes result in minimal redundancy via exploiting the underlying physics, they result in non-negligible complexity increase and access speed limitation since pages cannot be read separately. In this paper, we suggest new constrained coding schemes that have low-complexity and preserve the desirable high access speed in modern Flash devices. The idea is to eliminate error-prone patterns by coding data either only on the left-most page (binary coding) or only on the two left-most pages ($4$-ary coding) while leaving data on all the remaining pages uncoded. Our coding schemes work for any number of levels $q \geq 4$ per cell, offer systematic encoding and decoding, and are capacity-approaching. Since the proposed schemes enable the separation of pages, except the two left-most pages in the case of $4$-ary coding, we refer to them as \textit{read-and-run (RR)} constrained coding schemes as opposed to schemes adopting \textit{read-and-wait} for other pages. The $4$-ary RR coding scheme is introduced in order to limit the rate loss incurred by the binary RR coding schemes, and we show that our $4$-ary RR coding scheme is also competitive when it comes to complexity and error propagation. We analyze the new RR coding schemes and discuss their impact on the probability of occurrence of different charge levels. We also demonstrate the performance improvement achieved via RR coding on a practical triple-level cell Flash device.

\end{abstract}


\section{Introduction}\label{sec_intro}

The history of constrained coding dates back to 1948, when Shannon represented a constrained sequence via a finite-state transition diagram (FSTD) and derived the capacity under a constraint \cite{shan_const}. Run-length-limited (RLL) codes were introduced by Tang and Bahl in 1970 to support the evolution of magnetic recording at that time \cite{tang_bahl}, and these codes were based on lexicographic indexing. In 1973, Cover presented a result about enumerative coding \cite{cover_lex} that will prove fundamental for the design of constrained codes based on lexicographic indexing decades later. Among other researchers, Franaszek developed constrained codes based on finite-state machines (FSMs) derived from FSTDs \cite{franaszek}. In 1983, Adler, Coppersmith, and Hassner introduced a systematic method to develop constrained codes based on FSMs \cite{ach_fsm}. Details about the history of constrained coding until 1998 are in \cite{immink_surv}.

Because of their ability to improve performance via eliminating error-prone data patterns and undesirable sequences, constrained codes have a plethora of applications. They find application in one-dimensional (1D) magnetic recording devices, both the old ones, which are based on peak detection, and the modern ones, which are based on sequence detection \cite{vasic_prc, ahh_loco}. They can also be combined with robust signal detection using machine learning \cite{zheng_prnn}. They find application in the emerging two-dimensional (2D) magnetic recording devices as well \cite{wood_tdmr, bd_tdmr}. Moreover, constrained codes are used to achieve DC balance and self-calibration in optical recording devices \cite{immink_opt} in addition to many computer standards for data transmission \cite{saade_comp}.

In Flash devices, charge propagation from cells programmed to high charge levels into cells programmed to lower charge levels is the main reason behind inter-cell interference (ICI) \cite{lee_ici}. This is correct for any number $q$ of charge levels per cell. Mitigating ICI results in remarkable lifetime gains in Flash as demonstrated in \cite{veeresh_mlc} for multi-level cell (MLC) Flash ($q=4$). There are data patterns that are considered usual suspects for contributing most to ICI. Coding to eliminate data patterns resulting in consecutive levels $(q-1)0(q-1)$ was considered in \cite{ravi_const} and \cite{chee_ici}. Coding to eliminate data patterns resulting in consecutive levels $(q-1)\mu(q-1)$, also called level patterns, for all $\mu < q-1$, was presented in \cite{veeresh_mlc}, \cite{chee_ici}, and \cite{ahh_qaloco}.

A number of recent results revisited \cite{tang_bahl} and \cite{cover_lex} in order to produce efficient constrained codes based on lexicographic indexing, and one example is \cite{braun_lex}. Another example is \cite{ahh_loco}, in which we introduced binary symmetric lexicographically-ordered constrained (S-LOCO) codes and demonstrated density gains in a modern 1D magnetic recording system. We extended our result to single-level cell (SLC) Flash memories ($q=2$) \cite{ahh_aloco} then to Flash memories with any number $q$ of levels per cell \cite{ahh_qaloco}. Moreover, we devised a general method to design LOCO codes for any finite set of patterns to forbid \cite{ahh_general}, which will be useful in this paper. We studied the power spectra of binary LOCO codes in \cite{jes_psd}. LOCO codes are capacity-achieving, simple, and easily reconfigurable \cite{ahh_qaloco, ahh_general}.

While the constrained codes in \cite{chee_ici} and \cite{ahh_qaloco} are quite efficient in terms of rate, they require all Flash pages to be processed together, which negatively affects the access speed. In this paper, we propose binary \textit{read-and-run (RR)} constrained coding schemes that allow pages to be accessed separately in modern Flash devices, thus preserving high access speed. Our binary RR coding schemes incur small rate loss and work for any Flash device with $q \geq 4$ levels per cell. The key idea is that the constrained code is applied only on one page, the left-most page, while no coding is applied on the other $\log_2 q-1$ pages. We present a 2D RR coding scheme as well as a 1D RR coding scheme that is based on LOCO codes, and we name the latter binary RR-LOCO coding. Furthermore, we present a 1D $4$-ary RR coding scheme that is based on LOCO codes, which we name $4$-ary RR-LOCO coding, in order to further reduce the rate loss without impacting the device reliability. In particular, we apply constrained coding on two pages, the two left-most pages, while no coding is applied on the other $\log_2 q-2$ pages. Therefore, all pages are separated except the two left-most ones. Our $4$-ary RR coding scheme works for any Flash device with $q \geq 8$ levels per cell.\footnote{This $4$-ary RR coding scheme works for $q=4$ as well, but with more benign patterns forbidden and with no page separation.} We show that our $4$-ary RR coding scheme can even outperform the binary RR coding schemes at capacity-approaching rates in terms of both complexity and error propagation. There are techniques in the literature that allow page separation; however, they are either incurring notable rate loss \cite{veeresh_mlc} or designed for a specific Flash setup \cite{ravi_const}. We study various aspects about the proposed RR coding schemes, including charge-level probabilities. We introduce experimental results in a practical triple-level cell (TLC) Flash device ($q=8$) that demonstrate notable lifetime gains achieved by our coding schemes.

The rest of the paper is organized as follows. In Section~\ref{sec_map2d}, we discuss the detrimental patterns, the Flash mapping, and our 2D binary RR coding scheme. In Section~\ref{sec_rrloco2}, we introduce our 1D binary RR-LOCO coding scheme. In Section~\ref{sec_rrloco4}, we propose our 1D $4$-ary RR-LOCO coding scheme. In Section~\ref{sec_compropag}, we study the rate, complexity, and error propagation of the new schemes and make comparisons. In Section~\ref{sec_resultstlc}, we present the experimental results on TLC Flash. In Section~\ref{sec_concl}, we conclude the paper.

\section{Patterns, Mapping, and 2D RR Coding}\label{sec_map2d}

As implied in the introduction, literature works do not strictly agree on the set of forbidden patterns to operate on. Additionally, as the Flash device ages, the set of error-prone patterns is expected to expand \cite{ahh_qaloco}. According to our recent experimental tests and a machine learning-based ICI characterization \cite{sizheng_TLCGAN} of TLC Flash memories, we decided to focus on the set characterized as follows. Let
\begin{equation}
\beta_1, \overline{\beta}_1 \in \mathcal{V}_0 \triangleq \left \{\frac{q}{2}, \frac{q}{2}+1, \dots, q-1 \right \},
\end{equation}
where $q$ is the number of levels per Flash cell (a positive power of $2$) and $\mathcal{V}_1 = \{0, 1, \dots, q-1\} \setminus \mathcal{V}_0$. Then, the set of interest is the set resulting in the high-low-high level patterns in $\mathcal{L}_q$:\footnote{Levels are defined through their indices $\{0, 1, \dots, q-1\}$ for simplicity.}
\begin{equation}\label{eqn_forb}
\mathcal{L}_q \triangleq \{\beta_1\mu\overline{\beta}_1, \forall \beta_1,\overline{\beta}_1 \text{ } \vert \text{ } 0 \leq \mu < \min(\beta_1,\overline{\beta}_1)\}.
\end{equation}
This set already subsumes all $3$-tuple forbidden patterns adopted in the literature for Flash. This set can be relaxed by removing few patterns that have minimal impact on performance as we shall see in Section~\ref{sec_rrloco4}. A block inside the Flash device can be seen as a 2D grid of wordlines and bitlines, with a cell being placed at each intersection \cite{veeresh_mlc}. Level patterns in $\mathcal{L}_q$ are detrimental whether they occur on $3$ adjacent cells along the same wordline or along the same bitline.

\begin{example}
Consider an MLC Flash device, i.e., $q=4$. In this case, we have $\beta_1, \overline{\beta}_1 \in \{2,3\}$. Then, the set of interest is the set resulting in:
\begin{equation}
\mathcal{L}_4 = \{202, 212, 203, 213, 302, 312, 303, 313, 323\}.
\end{equation}
The last three elements in $\mathcal{L}_4$ are quite known \cite{veeresh_mlc, ravi_const, ahh_qaloco}.
\end{example}

\begin{algorithm}
\caption{Recursive Alternate Gray Mapping}
\begin{algorithmic}[1]
\State \textbf{Input:} Number of levels per cell $q$, and $p=\log_2 q$.
\State Define $\mathrm{map}$, a binary array of dimensions $q \times p$.
\State Set $\mathrm{map}(0,:) = \bold{1}^p$. \textit{(a sequence of $p$ $1$'s)}
\State \textbf{for} $i \in \{0, 1, \dots, p-1\}$ \textbf{do}
\State \hspace{2ex} \textbf{for} $j \in \{0, 1, \dots, 2^i-1\}$ \textbf{do}
\State \hspace{4ex} $\mathrm{map}(2^i+j,:) = \mathrm{map}(2^i-1-j,:)$.
\State \hspace{4ex} Flip the bit $\mathrm{map}(2^i+j,i)$. \textit{(each sequence in $\mathrm{map}$ is indexed from right to left by $0, 1, \dots, p-1$)}
\State \hspace{2ex} \textbf{end for}
\State \textbf{end for}
\State \textbf{Output:} Array $\mathrm{map}$ that maps each index to binary data.
\end{algorithmic}
\label{alg_ragm}
\end{algorithm}

Next, we discuss how to map from data to charge levels in Flash and vice versa. Since we are interested in page separation throughout this work, the mapping here is from a charge level out of $q$ possible ones to $\log_2 q$ binary bits, one for each page, and vice versa. Gray mapping offers the advantage that there is only one-bit difference between any two adjacent charge levels, which is valuable for error performance. We adopt a recursive alternate Gray mapping (RAGM), and Algorithm~\ref{alg_ragm} shows how to produce it for any $q \geq 4$. We highlight that RAGM has already been used in the literature in MLC Flash \cite{veeresh_mlc} and TLC Flash \cite{ravi_const}. Thus, RAGM is not strictly a new contribution.

\begin{example}
Consider a TLC Flash device, i.e., $q=8$. In this case, the output of Algorithm~\ref{alg_ragm}, which is RAGM, becomes:
\vspace{-0.5em}\begin{align}\label{eqn_map}
0&\longleftrightarrow111, \hspace{+3.0em} 1\longleftrightarrow110, \nonumber \\
2&\longleftrightarrow100, \hspace{+3.0em} 3\longleftrightarrow101, \nonumber \\
4&\longleftrightarrow001, \hspace{+3.0em} 5\longleftrightarrow000, \nonumber \\
6&\longleftrightarrow010, \hspace{+3.0em} 7\longleftrightarrow011.
\end{align}
\end{example}

Now, we are ready to discuss binary coding schemes. Let us first index the Flash pages the same way the bits in each sequence in the array $\mathrm{map}$ are indexed (see Algorithm~\ref{alg_ragm}). This means that the left-most page is the one indexed by $p-1$. From (\ref{eqn_forb}) and Algorithm~\ref{alg_ragm}, the level patterns in $\mathcal{L}_q$ correspond to binary patterns where the left-most page (pages) always has~(have) two $0$'s separated by some bit, i.e., $0x0$. Based on that, forbidding $\{000,010\}$ on the left-most page (pages) guarantees that no level pattern in $\mathcal{L}_q$ would appear while writing to a Flash device, with any $q \geq 4$, at least in the wordline (bitline) direction. This corresponds to an interleaved RLL $(d,k)=(0,1)$ constraint~\cite{siegel_nasit2015}. Notably, no coding on any other page is needed. Data will therefore be read from each page independently, and immediately passed to the low-density parity-check (LDPC) decoder to start its processing. This idea is the key idea of our binary RR constrained coding schemes.{\footnote{An equivalent scheme was proposed for MLC Flash, i.e., $q=4$, in \cite{siegel_nasit2015}.}
 
RR coding can be performed in the wordline direction only (1D), the bitline direction only (1D), or both directions (2D). Observe that binary RR coding will also prevent some benign level patterns, e.g., $474$, $555$, and $676$ in TLC Flash, resulting in inevitable rate loss. However, as we shall see in Section~\ref{sec_compropag}, this rate loss is small. Furthermore, some of these benign level patterns will be allowed when we shift from binary to $4$-ary coding, which reduces this rate loss, as we shall see in Section~\ref{sec_rrloco4}. RR-LOCO codes are capacity-approaching codes.

We start here with our scheme for 2D binary RR constrained coding. As the name suggests, we want to prevent the patterns in $\mathcal{R}^2=\{000,010\}$ from appearing at the left-most pages in both wordline and bitline directions in the Flash device through simple encoding and decoding. The encoding follows the rules:
\begin{enumerate}
\item On wordlines with indices congruent to $0$ or $1$ (mod $4$), you are allowed to write $0$'s and $1$'s freely in bit positions congruent to $0$ or $1$ (mod $4$) at the left-most pages.
\item On wordlines with indices congruent to $2$ or $3$ (mod $4$), you are allowed to write $0$'s and $1$'s freely in bit positions congruent to $2$ or $3$ (mod $4$) at the left-most pages.
\item In the other bit positions, you can only write $1$'s on wordlines at the left-most pages.
\end{enumerate}

This 2D binary RR constrained coding scheme is depicted in Fig.~\ref{fig_1}. It is clear from the figure that the patterns in $\mathcal{R}^2=\{000,010\}$ are eliminated on the left-most pages, which forbids all level patterns in $\mathcal{L}_q$, in both directions. Upon encoding, input data bits are freely placed at the positions marked by $x$ for the left-most pages, and they are directly placed (uncoded) at the other pages. Upon decoding, information at the positions marked by $1$ is omitted, and data bits at the remaining positions are read with no additional processing and with no correlation between different Flash pages.{\footnote{An equivalent 2D scheme forbidding patterns $\{101, 111\}$ on the right-most pages in both wordline and bitline directions in MLC Flash was proposed in~\cite{siegel_nasit2015}.}

This 2D binary scheme is ideal in terms of complexity, access speed, and error propagation (see Section~\ref{sec_compropag}). It might also seem notably better than any 1D scheme (binary or $4$-ary) in terms of performance. However, 1D schemes can achieve almost the same performance with higher rates, which we will discuss in more detail later.

\begin{figure}
\vspace{-0.5em}
\center
\includegraphics[trim={1.35in 2.2in 1.4in 1.0in}, width=4.0in]{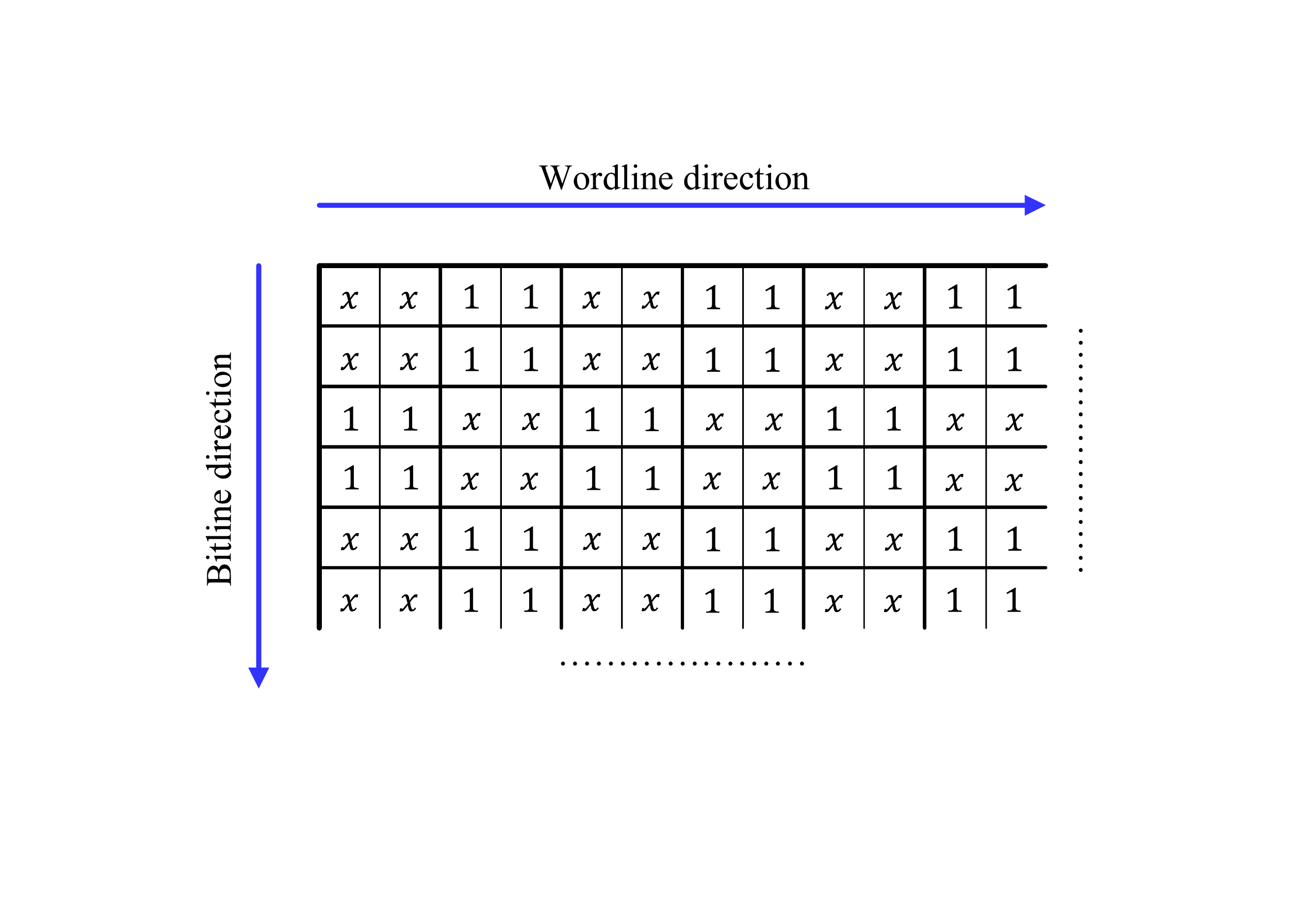}
\vspace{-0.0em}
\caption{The left-most pages of a 2D Flash grid with data encoded via the proposed 2D binary RR coding scheme. Symbol $x$ means bit can be $0$ or $1$ freely.}
\label{fig_1}
\vspace{-1.0em}
\end{figure}

\section{RR-LOCO Coding Over GF$(2)$}\label{sec_rrloco2}

In this section, we introduce a binary RR coding scheme that forbids $\{000,010\}$ on the left-most pages in either the wordline direction or the bitline direction, while leaving all other pages with no coding, which forbids the level patterns in $\mathcal{L}_q$ and achieves page separation. This scheme is the binary RR-LOCO coding scheme. The constrained code we apply is a binary LOCO code devised according to the general method in \cite{ahh_general}. We start by defining the proposed LOCO code.

\begin{definition}
A binary LOCO code $\mathcal{RC}^2_m$, where $m \geq 1$, that forbids the patterns in $\mathcal{R}^2=\{000,010\}$ is defined by the following properties:
\begin{enumerate}
\item Codewords in $\mathcal{RC}^2_m$ are defined over GF$(2) = \{0,1\}$ and are of length $m$ bits.
\item Codewords in $\mathcal{RC}^2_m$ are ordered lexicographically.
\item Codewords in $\mathcal{RC}^2_m$ do not have patterns in $\mathcal{R}^2$.
\item All codewords satisfying 1)--3) are included.
\end{enumerate}
\end{definition}

Lexicographic ordering is ordering codewords ascendingly according to the rule ``$0 < 1$'', where bit significance reduces from left to right \cite{tang_bahl, ahh_qaloco}. The first step to devise this binary LOCO code is to specify the group structure. Codewords in $\mathcal{RC}^2_m$, $m \geq 2$, can be partitioned into the following groups:
\begin{itemize}
\item Group~1: Codewords starting with $0011$ from the left.
\item Group~2: Codewords starting with $011$ from the left.
\item Group~3: Codewords starting with $1$ from the left.
\end{itemize}

The second step is to enumerate the codewords in $\mathcal{RC}^2_m$, which is done by Theorem~\ref{thm_enum}. Let $N_2(m) \triangleq \vert \mathcal{RC}^2_m \vert$.

\begin{theorem}\label{thm_enum}
The cardinality of a binary LOCO code $\mathcal{RC}^2_m$ is given by the recursive formula:
\begin{equation}\label{eqn_enum}
N_2(m) = N_2(m-1)+N_2(m-3)+N_2(m-4), \text{ } m \geq 2,
\end{equation}
where the defined cardinalities are:
\begin{equation}\label{eqn_def}
N_2(-3) \triangleq 0, \textup{ } N_2(-2) = N_2(-1) = N_2(0) \triangleq 1, \text{ and } N_2(1)=2.
\end{equation}
\end{theorem}

\begin{IEEEproof}
We compute the cardinalities of each group then add them all. Let the cardinality of Group~$i$ be $N_{2,i}$. As for Group~3 in $\mathcal{RC}^2_m$, there is a bijection between its codewords and the codewords in $\mathcal{RC}^2_{m-1}$ (attach $1$ from the left). Thus,
\begin{equation}\label{eqn_enum3}
N_{2,3}(m) = N_2(m-1).
\end{equation}
As for Group~2 in $\mathcal{RC}^2_m$, there is a bijection between its codewords and the codewords starting with $1$ from the left in $\mathcal{RC}^2_{m-2}$ (attach $01$ from the left). Thus using (\ref{eqn_enum3}),
\begin{equation}\label{eqn_enum2}
N_{2,2}(m) = N_{2,3}(m-2) = N_2(m-3).
\end{equation}
As for Group~1 in $\mathcal{RC}^2_m$, there is a bijection between its codewords and the codewords starting with $1$ from the left in $\mathcal{RC}^2_{m-3}$ (attach $001$ from the left). Thus using (\ref{eqn_enum3}),
\begin{equation}\label{eqn_enum1}
N_{2,1}(m) = N_{2,3}(m-3) = N_2(m-4).
\end{equation}
Adding (\ref{eqn_enum3}), (\ref{eqn_enum2}), and (\ref{eqn_enum1}) gives (\ref{eqn_enum}). The defined cardinalities, other than $N_2(1)$, can be computed by observing that $N_2(1)=2$, $N_2(2)=4$, $N_2(3)=6$, and $N_2(4)=9$, which sets up four equations. This observation is immediate given the forbidden patterns.
\end{IEEEproof}

Define a codeword $\bold{c}$ in $\mathcal{RC}^2_m$ as $\bold{c} \triangleq c_{m-1} c_{m-2} \dots c_0$, with $c_i \triangleq \zeta$ for $i \geq m$, where $\zeta$ represents ``out of codeword bounds''. The integer equivalent of a LOCO codeword bit $c_i$, $0 \leq i \leq m-1$, is $a_i$, i.e., $a_i$ is $0$ ($1$) when $c_i$ is $0$ ($1$). Denote the lexicographic index of a codeword $\bold{c}$ among all codewords in the LOCO code $\mathcal{RC}^2_m$ by $g_2(m,\bold{c})$, which is abbreviated to $g(\bold{c})$. In general, $g(\bold{c})$ is in $\{0, 1, \dots, N_2(m)-1\}$.

The third step is to specify the special cases of occurence for a $1$ inside a codeword in $\mathcal{RC}^2_m$. These cases are:
\begin{itemize}
\item Case 2: $c_{i+2}c_{i+1}c_i = 001$.
\item Case 3: $c_{i+2}c_{i+1}c_i = 011$.
\item Case 4: $c_{i+2}c_{i+1}c_i = 101$ or $c_{i+2}c_{i+1}c_i = \zeta 01$.
\end{itemize}
The typical or default case, Case 1, is simply the case of ``otherwise''. In particular, it is the case that $c_{i+2}c_{i+1}c_i=111$, $c_{i+2}c_{i+1}c_i=\zeta 11$, or $c_{i+1}c_i=\zeta 1$.

The fourth and fifth steps are to find the encoding-decoding rule, which specifies the mapping from index to codeword and vice versa. This rule for $\mathcal{RC}^2_m$ is given in Theorem~\ref{thm_rule}.

\begin{theorem}\label{thm_rule}
The relation between the lexicographic index $g(\bold{c})$, $\bold{c} \in \mathcal{RC}^2_m$, and the binary codeword $\bold{c}$ itself is given by:
\begin{align}\label{eqn_rule}
g(\bold{c}) = \sum_{i=0}^{m-1} a_i \Big [ (1-y_{i,1})N_2(i-2) + (1-y_{i,1}-y_{i,2})N_2(i-3) \Big ],
\end{align}
where $y_{i,1}$ and $y_{i,2}$ are specified as follows:
\begin{align}\label{eqn_rdef}
y_{i,1} &= 1 \text{ if } c_{i+2} c_{i+1} c_i \in \{001,011\}, \text{ and } y_{i,1} = 0 \text{ otherwise}, \nonumber \\
y_{i,2} &= 1 \text{ if } c_{i+1} c_i = 01 \textup{ s.t. } y_{i,1} = 0, \text{ and } y_{i,2} = 0 \text{ otherwise}.
\end{align}
\end{theorem}

\begin{IEEEproof}
We compute the contributions $g_{i,j}(c_i)$ of a bit $c_i$ under Case $j$, for all $j \in \{1,2,3,4\}$, in a binary LOCO codeword then merge them all. As for the typical case, which we index by $1$, this contribution is the number of codewords starting with $0$ from the left in $\mathcal{RC}^2_{i+1}$. Thus using (\ref{eqn_enum2}) and (\ref{eqn_enum1}),
\begin{align}\label{eqn_rule_1}
g_{i,1}(c_i) &= N_{2,2}(i+1) + N_{2,1}(i+1) \nonumber \\ &= N_2(i-2) + N_2(i-3).
\end{align}
As for Case 2 (Case 3), this contribution is the number of codewords starting with $000$ ($010$) from the left in $\mathcal{RC}^2_{i+3}$. Note that $000$ and $010$ are forbidden patterns. Thus,
\begin{align}\label{eqn_rule_23}
g_{i,2}(c_i) &= 0 \textup{ and } \nonumber \\ g_{i,3}(c_i) &= 0.
\end{align}
As for Case 4, this contribution is the number of codewords starting with $00$ from the left in $\mathcal{RC}^2_{i+2}$. Thus using (\ref{eqn_enum1}),
\begin{equation}\label{eqn_rule_4}
g_{i,4}(c_i) = N_{2,1}(i+2) = N_2(i-2).
\end{equation}
Using $y_{i,1}$ (for Cases 2 and 3) and $y_{i,2}$ (for Case 4) from (\ref{eqn_rdef}) along with $a_i$ to merge (\ref{eqn_rule_1}), (\ref{eqn_rule_23}), and (\ref{eqn_rule_4}) gives:
\begin{equation}\label{eqn_girule}
g_i(c_i) = a_i \Big [ (1-y_{i,1})N_2(i-2) + (1-y_{i,1}-y_{i,2})N_2(i-3) \Big ].
\end{equation}
Substituting (\ref{eqn_girule}) in $g(\bold{c}) = \sum_{i=0}^{m-1} g_i(c_i)$ gives (\ref{eqn_rule}).
\end{IEEEproof}

For brevity, we skip the sixth step, which is to assemble the encoding and decoding algorithms. These algorithms are a direct consequence of the rule in (\ref{eqn_rule}), and we refer the reader to \cite{tang_bahl}, \cite{ahh_qaloco}, \cite{ahh_general}, and \cite{laroia_const} for details. Note that we sometimes refer to $\mathcal{RC}^2_m$ as a \textit{1D binary RR-LOCO code}. The encoding-decoding rule of a LOCO code is the reason behind its low complexity algorithms, where reconfiguration becomes as easy as reprogramming an adder \cite{ahh_loco, ahh_general}.

\begin{remark}\label{rmk_kio}
If the coded bits are complemented before writing to pages, the set of forbidden patterns on the left-most pages becomes $\{101,111\}$ instead, which appears in \cite{veeresh_mlc} as well. In this case, the cardinality of the binary LOCO code remains as in (\ref{eqn_enum}), while the encoding-decoding rule becomes exactly that of a binary asymmetric LOCO code in \cite{ahh_aloco} for $x=1$:
\begin{equation}
g(\bold{c}) = \sum_{i=0}^{m-1} a_i N_2(i-a_{i+1}).\vspace{-0.1em}
\end{equation}
\end{remark}

Encoding and decoding on the left-most pages are just subtractions and additions. As for the remaining pages, data is written and read directly (uncoded). This guarantees simplicity and maintains high access speed via our 1D binary RR-LOCO coding~scheme.

\section{RR-LOCO Coding Over GF$(4)$}\label{sec_rrloco4}

In this section, we propose a 1D RR coding scheme over GF$(4)$, which is also based on LOCO codes. This scheme is our $4$-ary RR-LOCO coding scheme. The goal is to limit the rate loss resulting from binary RR coding schemes via coding on the two left-most pages. Finer classification of error-prone patterns, stemming from characterizing them via two bits instead of one, results in allowing some benign or less detrimental patterns, and therefore increasing the rate with negligible effect on performance.

We start by modifying the set of error-prone patterns. Let
\begin{align}\label{eqn_ws}
\theta_1, \overline{\theta}_1 \in \mathcal{W}_0 &\triangleq \left \{\frac{3q}{4}, \frac{3q}{4}+1, \dots, q-1 \right \}, \nonumber \\
\theta_2, \overline{\theta}_2 \in \mathcal{W}_1 &\triangleq \left \{\frac{q}{2}, \frac{q}{2}+1, \dots, \frac{3q}{4}-1 \right \}, \nonumber \\
\theta_3, \in \mathcal{W}_2 \cup \mathcal{W}_3, \textup{ } \mathcal{W}_2 &\triangleq \left \{\frac{q}{4}, \frac{q}{4}+1, \dots, \frac{q}{2}-1 \right \}, \textup{ } \mathcal{W}_3 \triangleq \left \{0, 1, \dots, \frac{q}{4}-1 \right \},
\end{align}
where $q$ is the number of levels per Flash cell (a positive power of $2$). While mathematically $q \geq 4$, we focus here on the case of $q \geq 8$. Then, the set of interest is the set resulting in the high-low-high level patterns in $\mathcal{L}'_q \subset \mathcal{L}_q$:
\begin{align}\label{eqn_forb4}
\mathcal{L}'_q \triangleq \{&\theta_1\eta\overline{\theta}_1, \forall \theta_1,\overline{\theta}_1 \text{ } \vert \text{ } 0 \leq \eta < \min(\theta_1,\overline{\theta}_1)\} \textup{ } \cup \textup{ }
\{\theta_1\theta_3\theta_2, \forall \theta_1,\theta_2,\theta_3\} \textup{ } \cup \nonumber \\
\{&\theta_2\theta_3\theta_1, \forall \theta_1,\theta_2,\theta_3\} \textup{ } \cup \textup{ }
\{\theta_2\theta_3\overline{\theta}_2, \forall \theta_2,\overline{\theta}_2,\theta_3\}.
\end{align}
This set also subsumes all $3$-tuple forbidden patterns adopted in the literature for Flash. The only difference between the set $\mathcal{L}'_q$ and the set $\mathcal{L}_q$ is that in the former, if either the left level is or the right level is or both levels are in $\mathcal{W}_1$, the middle level is always in $\mathcal{W}_2 \cup \mathcal{W}_3$. Our experimental results show that the level patterns in $\mathcal{L}_q \setminus \mathcal{L}'_q$ have very limited contribution to the errors occurring upon reading from the Flash device.

\begin{example}
Consider a TLC Flash device, i.e., $q=8$. In this case, we have $\theta_1, \overline{\theta}_1 \in \{6, 7\}$, $\theta_2, \overline{\theta}_2 \in \{4, 5\}$, and $\theta_3 \in \{0, 1, 2, 3\}$. Then, the difference between the two sets of interest is only one level pattern:
\begin{equation}
\mathcal{L}_{8} \setminus \mathcal{L}'_{8} = \{545\}.
\end{equation}
\end{example}

For mapping from charge levels to binary bits, we adopt the RAGM of Algorithm~\ref{alg_ragm}. Moreover, we index the Flash pages the same way the bits in each sequence in the array $\mathrm{map}$ are indexed using Algorithm~\ref{alg_ragm}. Therefore, we are interested here in the data on the two left-most pages indexed by $p-1$ and $p-2$. We adopt the following binary to $4$-ary mapping-demapping, where GF$(4) = \{0, 1, \alpha, \alpha^2\}$, for these two specific Flash pages:
\vspace{-0.1em}\begin{align}\label{eqn_map4}
11&\longleftrightarrow0 \textup{ } (\mathcal{W}_3), \hspace{+2.6em} 10\longleftrightarrow1 \textup{ } (\mathcal{W}_2), \nonumber \\
00&\longleftrightarrow\alpha \textup{ } (\mathcal{W}_1), \hspace{+2.5em} 01\longleftrightarrow\alpha^2 \textup{ } (\mathcal{W}_0).
\end{align}
The set of level patterns corresponding to each GF$(4)$ symbol is given between parenthesis.

We can see from (\ref{eqn_ws}), (\ref{eqn_forb4}), and (\ref{eqn_map4}) that the set of level patterns in $\mathcal{L}'_q$ can be forbidden in the wordline or the bitline direction by forbidding the $4$-ary patterns in the following set $\mathcal{R}^4$ from being written on the two left-most pages indexed by $p-1$ and $p-2$:
\vspace{-0.1em}\begin{align}\label{eqn_forbid4}
\mathcal{R}^4 = \{\alpha0\alpha, \alpha1\alpha, \alpha0\alpha^2, \alpha1\alpha^2, \alpha^20\alpha, \alpha^21\alpha, \alpha^20\alpha^2, \alpha^21\alpha^2, \alpha^2\alpha\alpha^2, \alpha^2\alpha^2\alpha^2\}.
\end{align}
Once again, no coding on any other page is needed. Data will therefore be read from each page independently, except the two left-most pages, and immediately passed to the low-density parity-check (LDPC) decoder to start its processing. This idea is the key idea of our $4$-ary RR constrained coding scheme.

Consider a TLC Flash device ($q=8$) once again. Forbidding the patterns in $\mathcal{R}^4$ on the two left-most pages instead of the patterns in $\mathcal{R}^2$ on the left-most page results in allowing many benign patterns that are forbidden if binary RR coding is adopted, e.g., $444$, $474$, and $555$.

Now, we introduce our $4$-ary RR coding scheme that forbids the patterns in $\mathcal{R}^4$ on the two left-most pages in either the wordline direction or the bitline direction, while leaving all other pages with no coding. The constrained code we apply is a $4$-ary LOCO code devised according to the general method in \cite{ahh_general}. We start by defining the proposed LOCO code.

\begin{definition}
A $4$-ary LOCO code $\mathcal{RC}^4_m$, where $m \geq 1$, that forbids the patterns in $\mathcal{R}^4$ is defined by the following properties:
\begin{enumerate}
\item Codewords in $\mathcal{RC}^4_m$ are defined over GF$(4) = \{0,1,\alpha,\alpha^2\}$ and are of length $m$ symbols.
\item Codewords in $\mathcal{RC}^4_m$ are ordered lexicographically.
\item Codewords in $\mathcal{RC}^4_m$ do not have patterns in $\mathcal{R}^4$.
\item All codewords satisfying 1)--3) are included.
\end{enumerate}
\end{definition}

Lexicographic ordering here is ordering codewords ascendingly according to the rule ``$0 < 1 < \alpha < \alpha^2$'', where symbol significance reduces from left to right \cite{tang_bahl, ahh_qaloco}. The first step to devise this $4$-ary LOCO code is to specify the group structure. Let $\gamma_1$ and $\gamma_2$ be in $\{0,1\}$. Codewords in $\mathcal{RC}^4_m$, $m \geq 3$, can be partitioned into the following groups:
\begin{itemize}
\item Group~1: Codewords starting with $\gamma_1$, $\forall \gamma_1$, from the left.
\item Group~2: Codewords starting with $\alpha \gamma_1 \gamma_2$, $\forall \gamma_1, \gamma_2$, from the left.
\item Group~3: Codewords starting with $\alpha \alpha$ or $\alpha \alpha^2$ from the left.
\item Group~4: Codewords starting with $\alpha^2 \gamma_1 \gamma_2$, $\forall \gamma_1, \gamma_2$, from the left.
\item Group~5: Codewords starting with $\alpha^2 \alpha \gamma_1 \gamma_2$, $\forall \gamma_1, \gamma_2$, from the left..
\item Group~6: Codewords starting with $\alpha^2 \alpha \alpha$ from the left.
\item Group~7: Codewords starting with $\alpha^2 \alpha^2 \gamma_1 \gamma_2$, $\forall \gamma_1, \gamma_2$, from the left..
\item Group~8: Codewords starting with $\alpha^2 \alpha^2 \alpha$ from the left.
\end{itemize}

The second step is to enumerate the codewords in $\mathcal{RC}^4_m$, which is done by Theorem~\ref{thm_enum4}. Let $N_4(m) \triangleq \vert \mathcal{RC}^4_m \vert$.

\begin{theorem}\label{thm_enum4}
The cardinality of a $4$-ary LOCO code $\mathcal{RC}^4_m$ is given by the recursive formula:
\begin{align}\label{eqn_enum4}
N_4(m) &= 3N_4(m-1) - 2N_4(m-2) + 9N_4(m-3) \nonumber \\ &\hspace{+1.0em}+ 7N_4(m-4) + 6N_4(m-5) + 4N_4(m-6), \text{ } m \geq 3,
\end{align}
where the defined cardinalities are:
\begin{align}\label{eqn_def4}
N_4(-5) \triangleq \frac{1}{32}, \textup{ } N_4(-4) \triangleq -\frac{1}{16}, \textup{ } N_4(-3) &\triangleq 0, \textup{ } N_4(-2) \triangleq \frac{1}{4}, \textup{ } N_4(-1) \triangleq \frac{1}{2}, \textup{ } N_4(0) \triangleq 1, \nonumber \\ \text{and } N_4(1)&=4, \textup{ } N_4(2)=16.
\end{align}
\end{theorem}

\begin{IEEEproof}
We compute the cardinalities of each group then add them all. Let the cardinality of Group~$i$ be $N_{4,i}$. As for Group~1 in $\mathcal{RC}^4_m$, there is a surjection between its codewords and the codewords in $\mathcal{RC}^4_{m-1}$ (attach $0$ or $1$ from the left). Thus,
\begin{equation}\label{eqn_enum4_1}
N_{4,1}(m) = 2N_4(m-1).
\end{equation}
As for Group~2 in $\mathcal{RC}^4_m$, there is a surjection between its codewords and the codewords in $\mathcal{RC}^4_{m-3}$. Thus,
\begin{equation}\label{eqn_enum4_2}
N_{4,2}(m) = (2)(2)N_4(m-3) = 4N_4(m-3).
\end{equation}
As for Group~3 in $\mathcal{RC}^4_m$, there is a bijection between its codewords and the codewords starting with $\alpha$ or $\alpha^2$ from the left in $\mathcal{RC}^4_{m-1}$. Thus using (\ref{eqn_enum4_1}),
\begin{equation}\label{eqn_enum4_3}
N_{4,3}(m) = N_4(m-1)-N_{4,1}(m-1) = N_4(m-1)-2N_4(m-2).
\end{equation}
As for Group~4 in $\mathcal{RC}^4_m$, the cardinality is the same as that of Group~2. Thus,
\begin{equation}\label{eqn_enum4_4}
N_{4,4}(m) = (2)(2)N_4(m-3) = 4N_4(m-3).
\end{equation}
As for Group~5 in $\mathcal{RC}^4_m$, it is handled in a way similar to that of Groups~2 and 4. Thus,
\begin{equation}\label{eqn_enum4_5}
N_{4,5}(m) = (2)(2)N_4(m-4) = 4N_4(m-4).
\end{equation}
As for Group~6 in $\mathcal{RC}^4_m$, there is a bijection between its codewords and the codewords starting with $\alpha$ from the left in $\mathcal{RC}^4_{m-2}$. Thus using (\ref{eqn_enum4_2}) and (\ref{eqn_enum4_3}),
\begin{equation}\label{eqn_enum4_6}
N_{4,6}(m) = N_{4,2}(m-2)+N_{4,3}(m-2) = N_4(m-3)-2N_4(m-4)+4N_4(m-5).
\end{equation}
As for Group~7 in $\mathcal{RC}^4_m$, the cardinality is the same as that of Group~5. Thus,
\begin{equation}\label{eqn_enum4_7}
N_{4,7}(m) = (2)(2)N_4(m-4) = 4N_4(m-4).
\end{equation}
As for Group~8 in $\mathcal{RC}^4_m$, there is a bijection between its codewords and the codewords starting with $\alpha^2 \alpha$ from the left in $\mathcal{RC}^4_{m-1}$. Thus using (\ref{eqn_enum4_5}) and (\ref{eqn_enum4_6}),
\begin{equation}\label{eqn_enum4_8}
N_{4,8}(m) = N_{4,5}(m-1)+N_{4,6}(m-1) = N_4(m-4)+2N_4(m-5)+4N_4(m-6).
\end{equation}

Adding (\ref{eqn_enum4_1}), (\ref{eqn_enum4_2}), (\ref{eqn_enum4_3}), (\ref{eqn_enum4_4}), (\ref{eqn_enum4_5}), (\ref{eqn_enum4_6}), (\ref{eqn_enum4_7}), and (\ref{eqn_enum4_8}) gives (\ref{eqn_enum4}). The defined cardinalities, other than $N_4(1)$ and $N_4(2)$, can be computed from the cardinalities at small values of $m$, which set up six equations.
\end{IEEEproof}

Define a codeword $\bold{c}$ in $\mathcal{RC}^4_m$ as $\bold{c} \triangleq c_{m-1} c_{m-2} \dots c_0$, with $c_i \triangleq \zeta$ for $i \geq m$, where $\zeta$ represents ``out of codeword bounds''. The integer equivalent of a LOCO codeword symbol $c_i$, $0 \leq i \leq m-1$, is $a_i$, i.e., $a_i$ is $0$, $1$, $2$, or $3$ when $c_i$ is $0$, $1$, $\alpha$, or $\alpha^2$, respectively. Denote the lexicographic index of a codeword $\bold{c}$ among all codewords in the LOCO code $\mathcal{RC}^4_m$ by $g_4(m,\bold{c})$, which is abbreviated to $g(\bold{c})$. In general, $g(\bold{c})$ is in $\{0, 1, \dots, N_4(m)-1\}$.

The third step is to specify the typical/special cases of occurence for a symbol in GF$(4) \setminus \{0\}$ inside a codeword in $\mathcal{RC}^4_m$. Let $\gamma$ be in $\{\zeta, 0, 1\}$ and $\chi$ be in $\{\alpha, \alpha^2\}$. These cases are:
\begin{itemize}
\item Case 1.a: $c_{i+1}c_i = \gamma 1$ or $c_{i+1}c_i = \gamma \alpha$, for all $\gamma$.
\item Case 1.b: $c_{i+1}c_i = \gamma \alpha^2$, for all $\gamma$.
\item Case 2: $c_{i+1}c_i = \chi 1$ or $c_{i+1}c_i = \chi \alpha$, for all $\chi$.
\item Case 3: $c_{i+1}c_i = \alpha \alpha^2$.
\item Case 4: $c_{i+1}c_i = \alpha^2 \alpha^2$.
\end{itemize}
The typical or default case is Case 1 (Case 1.a and Case 1.b combined).

The fourth and fifth steps are to find the encoding-decoding rule, which specifies the mapping from index to codeword and vice versa. This rule for $\mathcal{RC}^4_m$ is given in Theorem~\ref{thm_rule4}.

\begin{theorem}\label{thm_rule4}
The relation between the lexicographic index $g(\bold{c})$, $\bold{c} \in \mathcal{RC}^4_m$, and the $4$-ary codeword $\bold{c}$ itself is given by:
\begin{align}\label{eqn_rule4}
g(\bold{c}) = \sum_{i=0}^{m-1} \Big [ &\big[ (y_{i,1} + y'_{i,1}) a_i + y_{i,3} \big] N_4(i) + \big[ 2(y_{i,2}a_i + y_{i,3} - y'_{i,1}) + 5y_{i,\textup{d}} \big] N_4(i-1) \nonumber \\
&\big[ 4(y'_{i,1} + y_{i,3}) + 2y_{i,\textup{d}} \big] N_4(i-2) + 4y_{i,\textup{d}} N_4(i-3) \Big ],
\end{align}
where $y_{i,1}$, $y'_{i,1}$, $y_{i,2}$, $y_{i,3}$, and $y_{i,\textup{d}}$ are specified as follows:
\begin{align}\label{eqn_rdef4}
y_{i,1} &= 1 \text{ if } c_{i+1} c_i \in \{\gamma 1,\gamma \alpha \textup{ } | \textup{ } \forall \gamma\}, \text{ and } y_{i,1} = 0 \text{ otherwise}, \nonumber \\
y'_{i,1} &= 1 \text{ if } c_{i+1} c_i \in \{\gamma \alpha^2 \textup{ } | \textup{ } \forall \gamma\}, \text{ and } y'_{i,1} = 0 \text{ otherwise}, \nonumber \\
y_{i,2} &= 1 \text{ if } c_{i+1} c_i \in \{\chi 1,\chi \alpha \textup{ } | \textup{ } \forall \chi\}, \text{ and } y_{i,2} = 0 \text{ otherwise}, \nonumber \\
y_{i,3} &= 1 \text{ if } c_{i+1} c_i = \alpha \alpha^2, \text{ and } y_{i,3} = 0 \text{ otherwise}, \nonumber \\
y_{i,\textup{d}} &= 1 \text{ if } c_{i+1} c_i = \alpha^2 \alpha^2, \text{ and } y_{i,\textup{d}} = 0 \text{ otherwise}.
\end{align}
\end{theorem}

\begin{IEEEproof}
We compute the contributions $g_{i,j}(c_i)$ of a symbol $c_i$ under Case $j$, for all $j$ in $\{1,2,3,4\}$, in a $4$-ary LOCO codeword then merge them all. As for the typical case, Situation a, which we index by $1.a$, this contribution is the number of codewords starting with $c'_i < c_i$, where $c_i \in \{1, \alpha\}$, from the left in $\mathcal{RC}^4_{i+1}$. Thus using (\ref{eqn_enum4_1}),
\begin{align}\label{eqn_rule4_1a}
g_{i,1.a}(c_i) &= a_i N_4(i+1-1) = a_i N_4(i).
\end{align}
As for the typical case, Situation b, which we index by $1.b$, this contribution is the number of codewords starting with $c'_i < c_i$, where $c_i = \alpha^2$, from the left in $\mathcal{RC}^4_{i+1}$. Thus using (\ref{eqn_enum4_1}), (\ref{eqn_enum4_2}), and (\ref{eqn_enum4_3}),
\begin{align}\label{eqn_rule4_1b}
g_{i,1.b}(c_i) &= N_{4,1}(i+1) + N_{4,2}(i+1) + N_{4,3}(i+1) \nonumber \\ &= 3N_4(i) - 2N_4(i-1) + 4N_4(i-2).
\end{align}
As for Case 2, this contribution is the number of codewords starting with $c'_i \gamma_1$, $c'_i < c_i$, where $c_i \in \{1, \alpha\}$ and $\gamma_1 \in \{0,1\}$, from the left in $\mathcal{RC}^4_{i+1}$. Thus using (\ref{eqn_enum4_1}),
\begin{align}\label{eqn_rule4_2}
g_{i,2}(c_i) &= a_i N_{4,1}(i) = 2a_i N_4(i-1).
\end{align}
As for Case 3, this contribution is the number of codewords starting with $\alpha c'_i$, $c'_i < c_i$, where $c_i = \alpha^2$, from the left in $\mathcal{RC}^4_{i+2}$. Those are all the codewords starting with $\gamma_1 \gamma_2$, for all $\gamma_1$ and $\gamma_2$, from the left in $\mathcal{RC}^4_{i+1}$ plus all the codewords starting with $\alpha$ from the left in $\mathcal{RC}^4_{i+1}$. Thus using (\ref{eqn_enum4_1}), (\ref{eqn_enum4_2}), and (\ref{eqn_enum4_3}),
\begin{align}\label{eqn_rule4_3}
g_{i,3}(c_i) &= 2N_{4,1}(i) + N_{4,2}(i+1) + N_{4,3}(i+1) \nonumber \\ &= N_4(i) + 2N_4(i-1) + 4N_4(i-2).
\end{align}
As for Case 4, this contribution is the number of codewords starting with $\alpha^2 c'_i$, $c'_i < c_i$, where $c_i = \alpha^2$, from the left in $\mathcal{RC}^4_{i+2}$. Those are all the codewords starting with $\gamma_1 \gamma_2$, for all $\gamma_1$ and $\gamma_2$, from the left in $\mathcal{RC}^4_{i+1}$ plus all the codewords starting with $\alpha^2 \alpha$ from the left in $\mathcal{RC}^4_{i+2}$. Thus using (\ref{eqn_enum4_1}), (\ref{eqn_enum4_5}), and (\ref{eqn_enum4_6}),
\begin{align}\label{eqn_rule4_4}
g_{i,4}(c_i) &= 2N_{4,1}(i) + N_{4,5}(i+2) + N_{4,6}(i+2) \nonumber \\ &= 5N_4(i-1) + 2N_4(i-2) + 4N_4(i-3).
\end{align}
We use $y_{i,1}$, $y'_{i,1}$ (for Case 1), $y_{i,2}$ (for Case 2), $y_{i,3}$ (for Case 3), and $y_{i,\textup{d}}$ (for Case 4) from (\ref{eqn_rdef4}) along with $a_i$ to merge (\ref{eqn_rule4_1a}), (\ref{eqn_rule4_1b}), (\ref{eqn_rule4_2}), (\ref{eqn_rule4_3}), and (\ref{eqn_rule4_4}). We adopt the following merging functions, where $f^\textup{mer}_\ell(\cdot)$ is associated with $N_4(i+1-\ell)$:
\begin{align}\label{eqn_mergers}
f^\textup{mer}_1(\cdot) &= (y_{i,1} + y'_{i,1}) a_i + y_{i,3}, \nonumber \\
f^\textup{mer}_2(\cdot) &= 2(y_{i,2}a_i + y_{i,3} - y'_{i,1}) + 5y_{i,\textup{d}}, \nonumber \\
f^\textup{mer}_3(\cdot) &= 4(y'_{i,1} + y_{i,3}) + 2y_{i,\textup{d}}, \nonumber \\
f^\textup{mer}_4(\cdot) &= 4y_{i,\textup{d}}.
\end{align}
Therefore, the general form of the symbol contribution $g_i(c_i)$ is:
\begin{equation}\label{eqn_girule4}
g_i(c_i) = \sum_{\ell=1}^4 f^\textup{mer}_\ell(\cdot) N_4(i+1-\ell).
\end{equation}
Substituting (\ref{eqn_mergers}) and (\ref{eqn_girule4}) in $g(\bold{c}) = \sum_{i=0}^{m-1} g_i(c_i)$ gives (\ref{eqn_rule4}).
\end{IEEEproof}

\begin{remark}
Observe that the number of linearly independent merging variables is always less than the number of final cases \cite{ahh_general}. Here, $y_{i,\textup{d}}$ is dependent on the other merging variables as it can be written as $y_{i,\textup{d}} = \mathbbm{1}(a_i)(1-y_{i,1}-y'_{i,1}-y_{i,2}-y_{i,3})$, where $\mathbbm{1}(a_i)=1$ if $a_i > 0$ and $\mathbbm{1}(a_i)=0$ if $a_i=0$.
\end{remark}

For brevity, we again skip the sixth step, which is to assemble the encoding and decoding algorithms. These algorithms are a direct consequence of the rule in (\ref{eqn_rule4}), and we refer the reader to \cite{tang_bahl}, \cite{ahh_qaloco}, \cite{ahh_general}, and \cite{laroia_const} for details. Note that we sometimes refer to $\mathcal{RC}^4_m$ as a \textit{1D $4$-ary RR-LOCO code}.

\section{Rate, Complexity, and Error Propagation}\label{sec_compropag}

We start by calculating asymptotic rates. Unfortunately, deriving the capacity for 2D constrained codes is known to be notoriously hard. Therefore, we will derive the capacity $C^{\textup{1D}}_{\mathcal{L}_q}$ only under the 1D constrained coding setup, which is already higher than the capacity under the 2D setup. Thus, $C^{\textup{1D}}_{\mathcal{L}_q}$ serves as a ceiling for the highest achievable rate in a device where patterns in $\mathcal{L}_q$ are forbidden at least in one direction. We will shortly show that 1D constrained coding suffices in terms of performance.

An FSTD of a sequence where level patterns in $\mathcal{L}_q$ are forbidden is shown in Fig.~\ref{fig_2}. Based on this FSTD, the general adjacency matrix is (vectors are row vectors):
\begin{equation}\label{eqn_adjmax}
\bold{A}_1=\left[
\begin{array}{c|c|c|c}
\frac{q}{2} & \bold{1}_{\frac{q}{2}} & 0 & \bold{0}_{\frac{q}{2}-1} \\ \hline \vspace{-0.9em} &  &  & \\
\bold{0}^\mathrm{T}_{\frac{q}{2}} & \bold{U}^1_{\frac{q}{2}} & \frac{q}{2} \bold{1}^\mathrm{T}_{\frac{q}{2}} & \makecell{\bold{0}_{\frac{q}{2}-1} \\ \hline \bold{L}^1_{\frac{q}{2}-1}} \\ &  &  & \vspace{-0.9em} \\ \hline
\frac{q}{2} & \bold{0}_{\frac{q}{2}} & 0 & \bold{0}_{\frac{q}{2}-1} \\ \hline \vspace{-0.9em} &  &  & \\ 
\bold{0}^\mathrm{T}_{\frac{q}{2}-1} & \begin{array}{c|c} \hspace{-0.5em} \bold{I}_{\frac{q}{2}-1} & \bold{0}^\mathrm{T}_{\frac{q}{2}-1 \hspace{-0.5em}} \end{array} & \frac{q}{2} \bold{1}^\mathrm{T}_{\frac{q}{2}-1} & \makecell{\bold{0}_{\frac{q}{2}-1} \\ \hline \begin{array}{c|c} \hspace{-0.5em} \bold{L}^1_{\frac{q}{2}-2} & \bold{0}^\mathrm{T}_{\frac{q}{2}-2 \hspace{-0.5em}} \end{array}} \vspace{-0.9em} \\ &  &  &
\end{array}\right],
\end{equation}
where $\bold{U}^1_{\delta}$ ($\bold{L}^1_{\delta}$) is an upper (lower) only-ones triangular matrix of size $\delta \times \delta$. Thus and from \cite{shan_const}, the normalized capacity of a 1D constrained code forbidding the level patterns in $\mathcal{L}_q$ is:
\begin{equation}
C^{\textup{1D}}_{\mathcal{L}_q} = \frac{\log_2 (\lambda_{\max}(\bold{A}_1))}{\log_2 q},
\end{equation}
where $\lambda_{\max}(\bold{A})$ is the maximum real positive eigenvalue of the matrix~$\bold{A}$.\footnote{For positive integers $a+b \leq q$, the set $H$ of the $a$ largest levels, and the set $L$ of the $b$ smallest levels in $\{0, 1, \ldots, q-1\}$, a formula for the (count-constrained) capacity of the constrained system forbidding all level patterns in $\{\beta_1 \beta_2 \overline{\beta}_1 \text{ } \vert \text{ } \beta_1,\overline{\beta}_1\in H, \beta_2\in L\}$ was derived in~\cite{kashyap_isit2019}.}

\begin{figure}
\vspace{-0.5em}
\center
\includegraphics[trim={0.8in 1.4in 1.4in 1.0in}, width=4.0in]{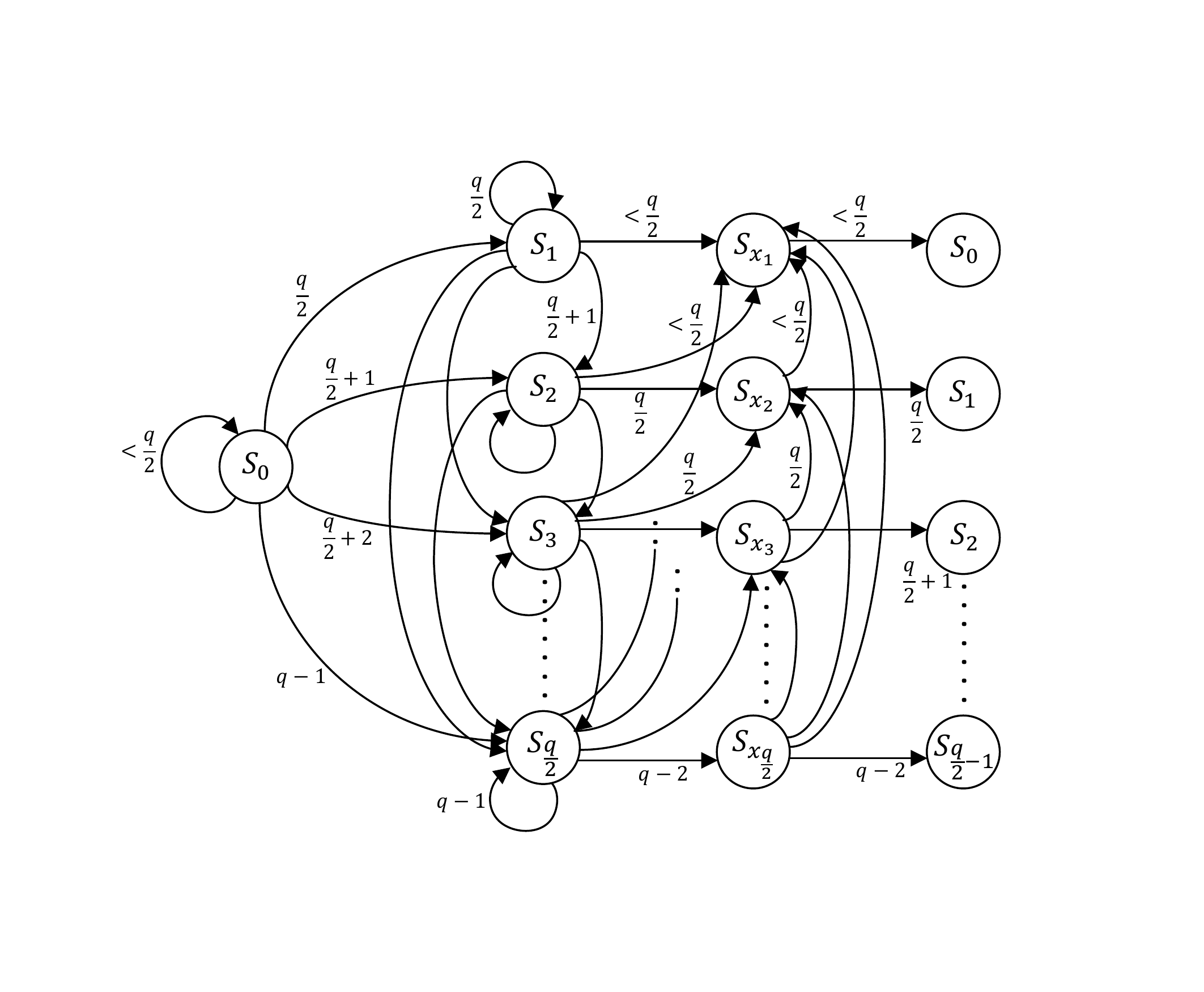}
\vspace{-0.0em}
\caption{An FSTD of a 1D constrained sequence forbidding level patterns in $\mathcal{L}_q$, for any $q$. Here, we operate directly on level patterns for simplicity.}
\label{fig_2}
\vspace{-0.3em}
\end{figure}

\begin{table}
\caption{Capacity Comparison Between $C^{\textup{1D}}_{\mathcal{L}_q}$, 1D Binary RR Capacity $C^{\textup{1D}}_{\textup{RR2}}$, and 1D $4$-ary RR Capacity $C^{\textup{1D}}_{\textup{RR4}}$}
\vspace{-0.5em}
\centering
\scalebox{1.00}
{
\begin{tabular}{|c|c|c|c|c|}
\hline
$q$ & \makecell{$C^{\textup{1D}}_{\mathcal{L}_q}$} & \makecell{$C^{\textup{1D}}_{\textup{RR2}}$} & \makecell{Capacity gap $\%$} & \makecell{$C^{\textup{1D}}_{\textup{RR4}}$}\\
\hline
$4$ & $0.8941$ & $0.8471$ & $5.257\%$ & $0.8859$ \\
\hline
$8$ & $0.9235$ & $0.8981$ & $2.750\%$ & $0.9239$ \\
\hline
$16$ & $0.9401$ & $0.9235$ & $1.766\%$ & $0.9429$ \\
\hline
$32$ & $0.9509$ & $0.9388$ & $1.272\%$ & $0.9544$ \\
\hline
\end{tabular}}
\label{table_1}
\vspace{-0.5em}
\end{table}

The capacity of a 2D binary code preventing $\{000,010\}$ is the capacity of a 2D $(0,1)$ RLL code, which is $\approx 0.5879$ \cite{kato_tcon}. Thus, the normalized capacity of our 2D RR coding scheme~is:
\begin{equation}
C^{\textup{2D}}_{\textup{RR2}} \approx \frac{0.5879+\log_2 q -1}{\log_2 q} = \frac{\log_2 q - 0.4121}{\log_2 q}.
\end{equation}

As mentioned above, the 1D constrained system where patterns in $\mathcal{R}^2=\{000,010\}$ are forbidden can be interpreted as an interleaved RLL $(d,k)=(0,1)$ constrained system, whose capacity is known to be $\log_2((1+\sqrt{5})/2)\approx 0.6942$. Thus, the normalized capacity of our 1D RR-LOCO coding scheme is:
\begin{equation}
C^{\textup{1D}}_{\textup{RR2}} = \frac{\log_2((1+\sqrt{5})/2)+\log_2 q -1}{\log_2 q} \approx \frac{\log_2 q - 0.3058}{\log_2 q}.
\end{equation}

The capacity gap between $C^{\textup{1D}}_{\mathcal{L}_q}$ and $C^{\textup{1D}}_{\textup{RR2}}$ for different values of $q$ is given in Table~\ref{table_1}. The table shows that the capacity gap is small, and it gets even smaller as $q$ increases.

The capacity $C^{\textup{1D}}_{\mathcal{L}'_q}$ of a 1D constrained system where the level patterns in $\mathcal{L}'_q$ are forbidden is slightly higher than $C^{\textup{1D}}_{\mathcal{L}_q}$ since $\mathcal{L}'_q \subset \mathcal{L}_q$. We skip the derivation of $C^{\textup{1D}}_{\mathcal{L}'_q}$ for brevity.

An FSTD of a 1D $4$-ary constrained system where patterns in $\mathcal{R}^4$ are forbidden is given in Fig.~\ref{fig_3}. The adjacency matrix is:
\begin{gather*}\label{eqn_adjacency4}
\bold{A}_2=
\begin{bmatrix}
2 & 1 & 1 & 0 & 0 & 0\vspace{-0.3em}\\
0 & 1 & 1 & 2 & 0 & 0\vspace{-0.3em}\\
0 & 0 & 0 & 2 & 1 & 1\vspace{-0.3em}\\
2 & 0 & 0 & 0 & 0 & 0\vspace{-0.3em}\\
0 & 1 & 0 & 2 & 0 & 0\vspace{-0.3em}\\
0 & 0 & 0 & 2 & 1 & 0
\end{bmatrix}.
\end{gather*}
The characteristic polynomial is:
\vspace{-0.1em}\begin{equation}\label{eqn_poly4}
\det(x \bold{I} - \bold{A}_2) = x^6 - 3x^5 + 2x^4 - 9x^3 - 7x^2 - 6x - 4.
\end{equation}
We can see that if $x$ is replaced by $\lambda_\textup{c} = \lambda_{\max}(\bold{A}_2)$, we get:
\begin{equation}\label{eqn_poly4_2}
\lambda^m_\textup{c} = 3\lambda^{m-1}_\textup{c} - 2\lambda^{m-2}_\textup{c} + 9\lambda^{m-3}_\textup{c} + 7\lambda^{m-4}_\textup{c} + 6\lambda^{m-5}_\textup{c} + 4\lambda^{m-6}_\textup{c},
\end{equation}
which is consistent with the cardinality recursion in (\ref{eqn_enum4}). The capacity of this $4$-ary constrained system is $\log_2 (\lambda_{\max}(\bold{A}_2)) = \log_2 (3.4147) = 1.7718$ bits$\slash$symbol. Thus, the normalized capacity of our 1D $4$-ary RR-LOCO coding scheme is:
\begin{equation}
C^{\textup{1D}}_{\textup{RR4}} = \frac{1.7718+\log_2 q -2}{\log_2 q} \approx \frac{\log_2 q - 0.2282}{\log_2 q}.
\end{equation}

Table~\ref{table_1} shows the capacity gain achieved by the 1D $4$-ary RR scheme over the 1D binary RR schemes, and we will show that the performance, i.e., the Flash device protection, is nearly the same. An interesting observation is that for $q \in \{8, 16, 32\}$, the capacity of our 1D $4$-ary RR scheme $C^{\textup{1D}}_{\textup{RR4}}$ is slightly higher than $C^{\textup{1D}}_{\mathcal{L}_q}$.

\begin{figure}
\vspace{-0.5em}
\center
\includegraphics[trim={1.1in 1.4in 1.4in 1.0in}, width=3.0in]{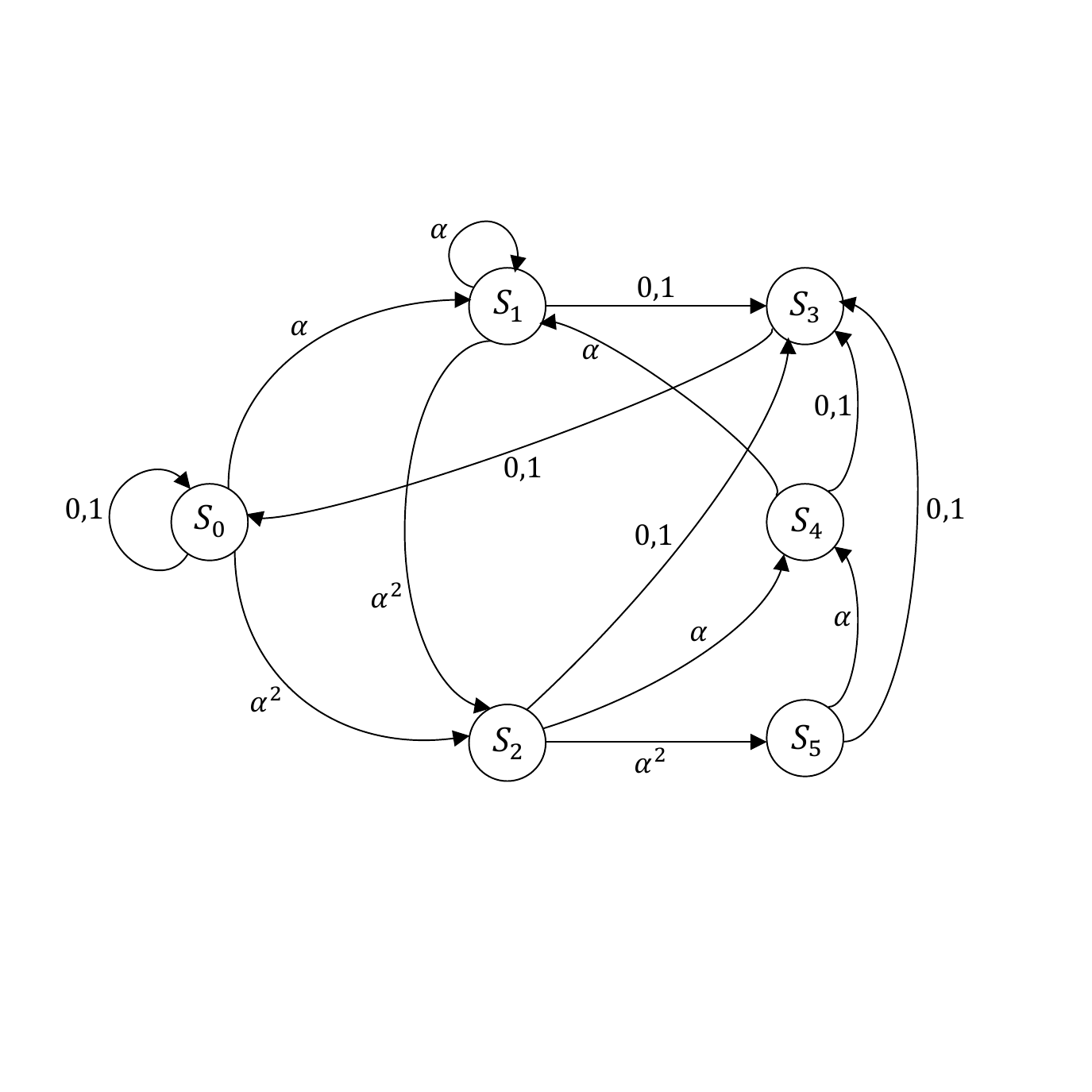}
\vspace{-1.0em}
\caption{An FSTD of a 1D $4$-ary constrained sequence forbidding patterns in $\mathcal{R}^4$.}
\label{fig_3}
\vspace{-0.5em}
\end{figure}

Next, we discuss the finite-length rates. First, the normalized rate of our 2D binary RR constrained coding scheme is:
\begin{equation}\label{eqn_rate2d}
R^{\textup{2D}}_{\textup{RR2}} = \frac{0.5+\log_2 q -1}{\log_2 q} = \frac{\log_2 - 0.5}{\log_2 q}
\end{equation}
since the rate of our left-most page coding is $0.5$.

Regarding our 1D binary RR-LOCO coding scheme, we bridge with the pattern $11$ between consecutive codewords in $\mathcal{RC}^2_m$ on the left-most page, and we remove the codeword $\bold{1}^m$ for self-clocking \cite{ahh_qaloco, ahh_general}. Thus, the rate on the left-most page is $\lfloor \log_2 (N_2(m)-1) \rfloor/(m+2)$, and the normalized rate of our 1D binary RR-LOCO coding scheme is:
\begin{equation}\label{eqn_rate1d}
R^{\textup{1D}}_{\textup{RR2}} = \frac{1}{\log_2 q} \left [ \frac{\lfloor \log_2 (N_2(m)-1) \rfloor}{m+2}+\log_2 q -1 \right ].
\end{equation}

1D binary RR-LOCO coding schemes are capacity-achieving schemes in the sense that the limit as $m \rightarrow \infty$ of $R^{\textup{1D}}_{\textup{RR2}}$ is $C^{\textup{1D}}_{\textup{RR2}}$ (see also \cite{ahh_qaloco}). Another capacity-achieving 1D RR constrained coding scheme, implementable using enumerative coding without the need for bridging bits, can be obtained by interleaving codewords from an optimal block code for the RLL $(d,k)=(0,1)$ constraint~\cite{marcus_jsac1992} on the left-most pages. LOCO codes, however, offer simplicity and reconfigurability, which is important as the device ages \cite{ahh_qaloco}.

Regarding our 1D $4$-ary RR-LOCO coding scheme, we cannot bridge with a single GF$(4)$ symbol between consecutive codewords in $\mathcal{RC}^4_m$ on the two left-most pages since any symbol separating $\alpha^2$ and $\alpha^2$ generates a forbidden pattern. We propose a novel two-symbol bridging in which \textit{we can encode input information bits within the bridging interval} as follows:
\begin{itemize}
\item For input information bits $00 \in$ GF$(2)$, bridge with $00 \in$ GF$(4)$.
\item For input information bits $01 \in$ GF$(2)$, bridge with $01 \in$ GF$(4)$.
\item For input information bits $10 \in$ GF$(2)$, bridge with $10 \in$ GF$(4)$.
\item For input information bits $11 \in$ GF$(2)$, bridge with $11 \in$ GF$(4)$.
\end{itemize}
While it has no effect on the asymptotic rate, this bridging scheme remarkably reduces the code length at which a specific rate is achieved, significantly reducing the complexity and error propagation in consequence.

To achieve self-clocking, we remove the two codewords $\bold{0}^m$ and $\bold{1}^m$, which is expected given the bridging above \cite{ahh_qaloco, ahh_general}. Thus, the rate on the two left-most pages is $(\lfloor \log_2 (N_4(m)-2) \rfloor + 2)/(m+2)$ bits$\slash$symbol, and the normalized rate of our 1D $4$-ary RR-LOCO coding scheme is:
\begin{equation}\label{eqn_rate1d4}
R^{\textup{1D}}_{\textup{RR4}} = \frac{1}{\log_2 q} \left [ \frac{\lfloor \log_2 (N_4(m)-2) \rfloor + 2}{m+2}+\log_2 q -2 \right ].
\end{equation}

1D $4$-ary RR-LOCO coding schemes are capacity-achieving schemes in the sense that the limit as $m \rightarrow \infty$ of $R^{\textup{1D}}_{\textup{RR4}}$ is $C^{\textup{1D}}_{\textup{RR4}}$. RR-LOCO codes offer simplicity and reconfigurability, which is important as the device ages \cite{ahh_qaloco}.

The 2D binary RR constrained coding scheme we propose requires no additional complexity for encoding and decoding since data is written/read directly to/from specific positions on the left-most page and directly to/from all positions on other pages. As for the 1D binary RR-LOCO coding scheme, the complexity is governed by the size of the adder that executes the encoding-decoding rule, which is:
\begin{equation}\label{eqn_msg}
s_{\textup{2}} = \lfloor \log_2 (N_2(m)-1) \rfloor
\end{equation}
bits. Similarly and as for the 1D $4$-ary RR-LOCO coding scheme, the complexity is governed by the adder size, which is:
\begin{equation}\label{eqn_msg4}
s_{\textup{4}} = \lfloor \log_2 (N_4(m)-2) \rfloor
\end{equation}
bits. For ease of implementation and to avoid affecting the access speed, we prefer to apply the 1D RR-LOCO coding schemes along wordlines instead of bitlines since the performance is very close, as demonstrated by the experimental results in Section~\ref{sec_resultstlc}.

Error propagation is the phenomenon that a single writing error results in multiple errors while reading. The 2D binary RR coding scheme does not incur any error propagation. Thus, the error propagation factor of it is $E^{\textup{2D}}_{\textup{RR2}}=1$. As for the 1D binary RR-LOCO coding scheme, there is no codeword-to-codeword error propagation. However, there exists limited error propagation resulting from the codeword-to-message conversion \cite{ahh_loco, ahh_qaloco} on the left-most page only. This error propagation reaches $s_{\textup{2}}/2$ bits on average, where $s_{\textup{2}}$ is the message length as well from (\ref{eqn_msg}). Consequently, the error propagation factor averaged over $\log_2 q$ pages is:
\begin{equation}
E^{\textup{1D}}_{\textup{RR2}} = \frac{1}{\log_2 q} \left [ \frac{s_{\textup{2}}}{2}+\log_2 q-1 \right ].
\end{equation}

As for the 1D $4$-ary RR-LOCO coding scheme, again there exists limited error propagation resulting solely from the LOCO codeword-to-message conversion \cite{ahh_loco, ahh_qaloco} on the two left-most pages. This error propagation reaches $s_{\textup{4}}/2$ bits on average, where $s_{\textup{4}}$ is the message length as well from (\ref{eqn_msg4}). Observe that there is no error propagation for the two additional bits encoded at each bridging interval to specify the two $4$-ary bridging symbols. Therefore, the average error propagation on any of these two left-most pages is:
\begin{equation}
\frac{s_{\textup{4}}}{2} \cdot \frac{m}{m+2} + 1 \cdot \frac{2}{m+2} = \frac{s_{\textup{4}}m + 4}{2(m+2)}.
\end{equation}
Consequently, the error propagation factor averaged over $\log_2 q$ pages is:
\begin{equation}
E^{\textup{1D}}_{\textup{RR4}} = \frac{1}{\log_2 q} \left [ 2 \cdot \frac{s_{\textup{4}}m + 4}{2(m+2)} +\log_2 q-2 \right ] = \frac{1}{\log_2 q} \left [ \frac{s_{\textup{4}}m + 4}{m+2} +\log_2 q-2 \right ].
\end{equation}

Another metric to compare 1D binary with 1D $4$-ary RR-LOCO coding schemes is the amount of coded data at a given rate. As this amount decreases, the code allows achieving the desired rate at a smaller length $m$, which is an advantage. Since for our 1D binary and 1D $4$-ary RR-LOCO coding schemes we use two bits and two symbols for bridging, respectively, these amounts of coded data,  $D^{\textup{1D}}_{\textup{RR2}}$ (binary) and $D^{\textup{1D}}_{\textup{RR4}}$ ($4$-ary) are:
\begin{align}
D^{\textup{1D}}_{\textup{RR2}} &= (m+2) \log_2 q, \textup{ $m$ is the length of } \mathcal{RC}^2_m,\\
D^{\textup{1D}}_{\textup{RR4}} &= (m+2) \log_2 q, \textup{ $m$ is the length of } \mathcal{RC}^4_m.
\end{align}

\begin{table}
\caption{Comparisons of Rate, Complexity, and Error Propagation at the Same Length Between 2D RR and 1D Binary RR Constrained Coding Schemes}
\vspace{-0.5em}
\centering
\scalebox{1.00}
{
\begin{tabular}{|c|c|c|c|c|c|c|}
\hline
$q$ & $m$ & $R^{\textup{2D}}_{\textup{RR2}}$ & $R^{\textup{1D}}_{\textup{RR2}}$ & $s_{\textup{2}}$ & $E^{\textup{2D}}_{\textup{RR2}}$ & $E^{\textup{1D}}_{\textup{RR2}}$ \\
\hline
$4$ & $7$ & $0.7500$ & $0.7778$ & $5$ & $1.000$ & $1.750$ \\
\hline
$4$ & $11$ & $0.7500$ & $0.8077$ & $8$ & $1.000$ & $2.500$ \\
\hline
$4$ & $21$ & $0.7500$ & $0.8261$ & $15$ & $1.000$ & $4.250$ \\
\hline
$8$ & $7$ & $0.8333$ & $0.8519$ & $5$ & $1.000$ & $1.500$ \\
\hline
$8$ & $11$ & $0.8333$ & $0.8718$ & $8$ & $1.000$ & $2.000$ \\
\hline
$8$ & $21$ & $0.8333$ & $0.8841$ & $15$ & $1.000$ & $3.167$ \\
\hline
$16$ & $7$ & $0.8750$ & $0.8889$ & $5$ & $1.000$ & $1.375$ \\
\hline
$16$ & $11$ & $0.8750$ & $0.9038$ & $8$ & $1.000$ & $1.750$\\
\hline
$16$ & $21$ & $0.8750$ & $0.9130$ & $15$ & $1.000$ & $2.625$ \\
\hline
\end{tabular}}
\label{table_2}
\vspace{-0.5em}
\end{table}

\begin{table}
\caption{Comparisons of Minimum Coded Data, Complexity, and Error Propagation to Achieve Certain Rate Between 1D Binary RR and 1D $4$-ary RR Constrained Coding Schemes}
\vspace{-0.5em}
\centering
\scalebox{1.00}
{
\begin{tabular}{|c|c|c|c|c|c|c|c|}
\hline
$q$ & Rate & $D^{\textup{1D}}_{\textup{RR2}}$ & $D^{\textup{1D}}_{\textup{RR4}}$ & $s_{\textup{2}}$ & $s_{\textup{4}}$ & $E^{\textup{1D}}_{\textup{RR2}}$ & $E^{\textup{1D}}_{\textup{RR4}}$ \\
\hline
$8$ & $0.8500$ & $27$ & $21$ & $5$ & $9$ & $1.500$ & $2.667$ \\
\hline
$8$ & $0.8750$ & $48$ & $24$ & $10$ & $11$ & $2.333$ & $3.250$ \\
\hline
$8$ & $0.8900$ & $138$ & $48$ & $31$ & $25$ & $5.833$ & $7.708$ \\
\hline
$8$ & $0.9000$ & $-$ & $60$ & $-$ & $32$ & $-$ & $10.000$ \\
\hline
$16$ & $0.8900$ & $48$ & $28$ & $7$ & $9$ & $1.625$ & $2.250$ \\
\hline
$16$ & $0.9050$ & $64$ & $32$ & $10$ & $11$ & $2.000$ & $2.688$\\
\hline
$16$ & $0.9150$ & $144$ & $48$ & $24$ & $18$ & $3.750$ & $4.333$ \\
\hline
$16$ & $0.9200$ & $288$ & $64$ & $49$ & $25$ & $6.875$ & $6.031$ \\
\hline
$16$ & $0.9300$ & $-$ & $100$ & $-$ & $41$ & $-$ & $9.970$ \\
\hline
\end{tabular}}
\label{table_3}
\vspace{-0.5em}
\end{table}

Table~\ref{table_2} gives the normalized rates, adder sizes, and error propagation factors of the proposed binary RR schemes under various parameters. The 1D binary RR-LOCO coding scheme has a remarkable rate~advantage that reaches $10.147\%$, $6.096\%$, and $4.343\%$ for $q=4$, $q=8$, and $q=16$, respectively, over the 2D binary RR constrained coding scheme. The 2D binary RR scheme has a clear advantage in terms of both complexity and error propagation as it requires no processing to encode and decode. Having said that, the error propagation factor of the 1D binary RR scheme decreases notably as $q$ increases. For example, $E^{\textup{1D}}_{\textup{RR2}}=2.625$ for $q=16$ and $m=21$, which is remarkably small given the code length.

In Table~\ref{table_3}, we compare 1D binary with 1D $4$-ary RR-LOCO coding schemes in a different way. In particular, we fix the normalized rate, and find the minimum amount of coded data and the minimum complexity (adder size) required to achieve this desired rate for the two coding schemes, in addition to the minimum error propagation associated with them.\footnote{Achieving a desired rate here means reaching a normalized rate greater than or equal to this desired rate.} The sign ``$-$'' is used in the table whenever the binary coding scheme cannot achieve such a rate. The main conclusions from Table~\ref{table_3} are:
\begin{itemize}
\item For $q=8$ and $q=16$, the $4$-ary coding scheme requires less coded data (smaller lengths) than the binary coding scheme does for all desired rates. The difference in favor of the $4$-ary coding scheme increases as the rate increases.

\item At lower rates, the complexity of the binary coding scheme is lower than that of the $4$-ary coding scheme. However, at rates $\geq 0.8900$ for $q=8$ and $\geq 0.9150$ for $q=16$, the $4$-ary coding scheme wins the complexity competition.
 
\item As expected, the binary coding scheme incurs less error propagation in general because LOCO coding is performed on one page only. However, at higher rates and higher $q$, the $4$-ary coding scheme becomes quite competitive to the intriguing extent that it already incurs less error propagation at rate $0.9200$ and $q=16$.
\end{itemize}

The 1D and 2D RR coding schemes can be used in the same device, but at different lifetime stages. A 1D RR-LOCO coding scheme, binary or $4$-ary, can be used when the device is relatively fresh or until a moderate number of program/erase (P/E) cycles, while the 2D RR constrained coding scheme can be used when the device ages, where preventing the error-prone patterns in both directions could make a difference and the associated rate loss could be acceptable. However, this performance difference is shown to be small in Section~\ref{sec_resultstlc}, at least for the TLC Flash device we used. The section also shows that the performance difference between 1D binary and 1D $4$-ary RR-LOCO coding schemes is negligible.

\begin{remark}
An idea that allows page separation for MLC Flash was introduced in \cite{veeresh_mlc}. However, the rate offered is only $0.7500$, which is significantly below the rates offered via our 1D binary RR coding scheme for MLC. Another idea that allows page separation for TLC Flash was introduced in \cite{ravi_const}. However, it only heuristically addresses the level pattern $707$. 
\end{remark}

\section{Experimental Results on TLC Flash}\label{sec_resultstlc}

To characterize the performance of the proposed RR constrained coding schemes, we conducted program/erase (P/E) cycling experiments on several blocks of a commercial 1X-nm TLC Flash chip, as follows:
\begin{enumerate}
\item Erase Flash memory block under test.
\item Program all pages of block under test with data. For uncoded experiments, program pseudo-random data at each P/E cycle. For RR experiments, program prepared data satisfying RR constraints at each P/E cycle. 
\item For each successive P/E cycle of RR experiments, ``rotate'' the data, so the data that was written on the page $i$ is written on the page $(i+1)$, wrapping around the last page to the first page. 
\item Record bit errors and compute channel bit error rate (BER) every $100$ P/E cycles.
\end{enumerate}
The PE cycling experiments were performed at room temperature in a continuous manner with no wait time between the erase-program-read operations.

Gray mappings used in Flash devices may vary between manufacturers and product generations. In our preliminary work~\cite{ahh_rrconst}, we modified the  forbidden  binary patterns in accordance with the device mapping so that RR coding on one page per wordline  would eliminate  most of the patterns in $\mathcal{L}_q$ that induce the most severe ICI (see Remark~\ref{rmk_kio}).

In this work, the $8$-ary encoded level sequences generated by the RR encoders described herein using the RAGM mapping were translated according to the device Gray mapping into the corresponding binary sequences for the lower, middle, and upper pages in the TLC Flash memory. Thus, the $8$-ary level sequences stored in the memory are precisely the RR-encoded level sequences (each cell is programmed to a level in $\{0, 1, \dots, q-1\}$).   

\begin{figure}
\vspace{-5.0em}
\center\includegraphics[trim={0.0in 0.0in 0.0in 0.0in}, width=6.5in]{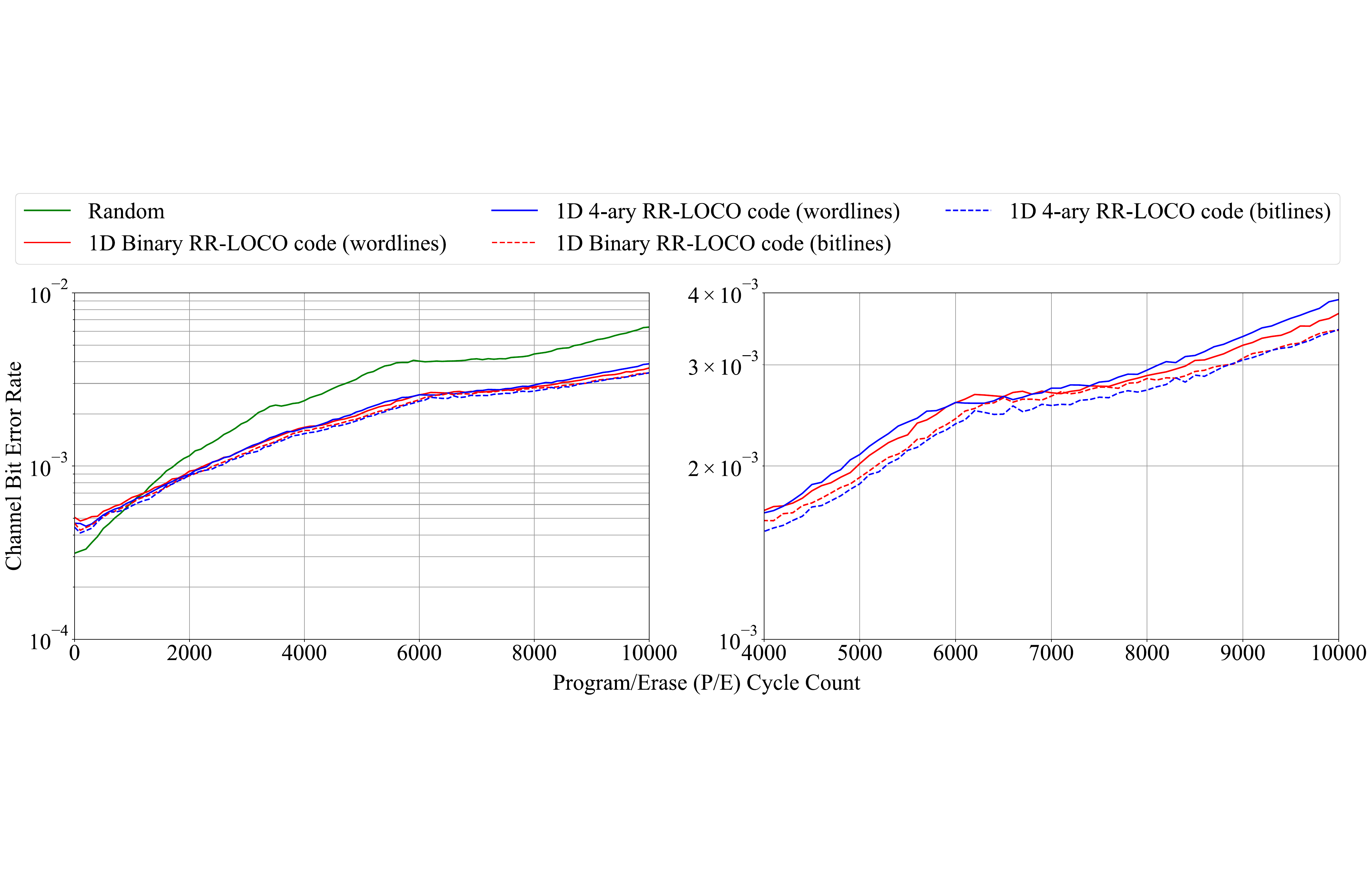}\vspace{-5.0em}
\caption{(Left) Measured average channel BER comparison when all pages are programmed with random data, 1D binary RR-LOCO coded data along wordlines or bitlines, and 1D $4$-ary RR-LOCO coded data along wordlines or bitlines  from P/E cycle $0$ to P/E cycle $10{,}000$. (Right) Measured average channel BER excluding random data from P/E cycle $4{,}000$ to P/E cycle $10{,}000$.}
\vspace{-0.5em}
\label{exp_RR_all_life}
\end{figure}

\begin{figure}
\center
\vspace{-2.5em}
\includegraphics[trim={0.0in 0.0in 0.0in 0.0in}, width=4.8in]{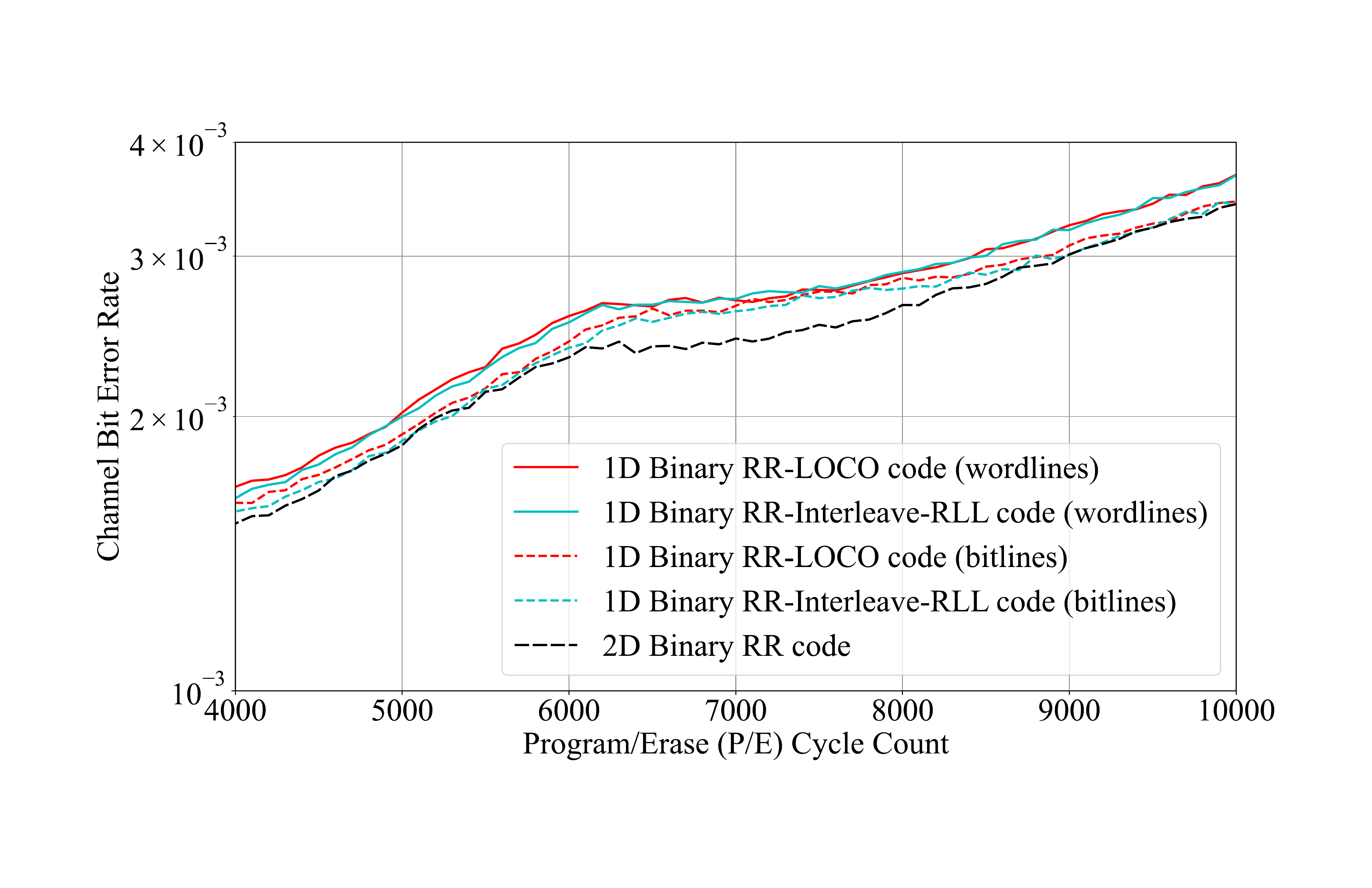}\vspace{-2.5em}
\caption{Measured average channel BER comparison of 1D binary RR-LOCO codes along wordlines or bitlines, 1D binary interleaved RLL-$(0,1)$ codes along wordlines or bitlines, and 2D binary RR code from P/E cycle $4{,}000$ to P/E cycle $10{,}000$.}
\vspace{-0.5em}
\label{exp_RR_high_PE}
\end{figure}

The left subfigure in Fig.~\ref{exp_RR_all_life} shows the channel BER from P/E cycle $0$ to P/E cycle $10{,}000$ using pseudo-random data, a rate $24{:}36$ 1D binary RR-LOCO code along wordlines or bitlines, and a rate $20{:}12$ bits/symbol 1D $4$-ary RR-LOCO code along wordlines or bitlines. The right subfigure in Fig.~\ref{exp_RR_all_life} shows the channel BER from P/E cycle $4{,}000$ to P/E cycle $10{,}000$ for these cases in more detail. Note that the binary RR code and $4$-ary RR code have the same overall rate: $R^{\textup{1D}}_{\textup{RR2}} = 8/9 \approx 0.8889$ using (\ref{eqn_rate1d}) and $R^{\textup{1D}}_{\textup{RR4}} = 8/9 \approx 0.8889$ using (\ref{eqn_rate1d4}). Therefore, the 1D binary coding scheme achieves about $99\%$ ($96\%$) of the capacity $C^{\textup{1D}}_{\textup{RR2}}$ ($C^{\textup{1D}}_{\mathcal{L}_q}$) and the 1D $4$-ary coding scheme achieves about $96\%$ of the capacity $C^{\textup{1D}}_{\textup{RR4}}$. 


As shown in Fig. \ref{exp_RR_all_life}, the uncoded performance is better than that of both binary and $4$-ary RR codes up to around $1{,}200$ P/E cycles and is notably worse thereafter. At the later stages of P/E cycling, ICI becomes severe and RR codes outperform the uncoded setting. Specifically, 1D binary RR-LOCO codes along wordlines increase device lifetime by about $1{,}800$ P/E cycles when channel BER is $2\times 10^{-3}$, representing a $57\%$ lifetime gain, and achieve about $3{,}700$ P/E cycles gain when channel BER is $3\times 10^{-3}$, corresponding to a $79\%$ lifetime gain. As shown in the right subfigure of Fig.~\ref{exp_RR_all_life}, the BER of 1D binary RR code along wordlines is almost the same as that of the $4$-ary RR code between $2{,}000$ and $8{,}000$ P/E cycles. When the P/E cycle count is larger than $8{,}000$, the BER of 1D binary RR code along wordlines is slightly better than that of the 1D $4$-ary RR code. In particular, when channel BER is $3\times 10^{-3}$, the 1D binary RR code along wordlines provides a lifetime that is about $300$ P/E cycles larger that than obtained with the 1D $4$-ary RR code along wordlines. Along the bitline direction, quite intriguingly, the performance of the 1D $4$-ary RR-LOCO code is generally better than, though close to, that of the 1D binary RR-LOCO code. The advantage of the 1D $4$-ary RR-LOCO is most pronounced from P/E cycle $6{,}300$ to P/E cycle $8{,}300$. 

Fig.~\ref{exp_RR_high_PE} compares the BER performance of different implementations of binary RR codes at high P/E cycles: the $24{:}36$ 1D binary RR-LOCO code along the wordline or bitline direction, the 1D binary interleaved $12{:}18$ RLL $(d,k)=(0,1)$ code (which has an overall block length $36$ after interleaving) along the wordline or bitline direction, and the 2D binary RR code. Using (\ref{eqn_rate2d}), we obtain $R^{\textup{2D}}_{\textup{RR2}} = 5/6 \approx 0.8333$. Therefore, the~2D coding scheme achieves about $93\%$ ($90\%$) of the capacity $C^{\textup{2D}}_{\textup{RR2}}$ ($C^{\textup{1D}}_{\mathcal{L}_q}$).

From Fig.~\ref{exp_RR_high_PE}, we have the following observations: the 1D RR coding schemes along the bitline direction achieve a slightly better performance than the 1D RR coding schemes along the wordline direction; the performance of the 2D RR constrained code is better than that of the 1D RR codes along any one direction; and the 1D RR coding schemes along the same direction have similar performance. For example, when channel BER is $2\times 10^{-3}$, the 2D binary RR coding increases lifetime by $100$ P/E cycles over the 1D binary RR-LOCO coding along bitlines and $300$ P/E cycles over the 1D binary RR-LOCO coding along wordlines. When channel BER is $3\times 10^{-3}$ and the wear condition of the Flash device is severe, the 2D binary RR coding outperforms the 1D binary RR-LOCO coding along bitlines by about $200$ P/E cycles and the 1D binary RR-LOCO coding along wordlines by about $600$ P/E cycles. 

These measurements confirm some of the claimed practical advantages of $4$-ary RR codes.  The performance results of the 1D binary RR-LOCO code and the 1D $4$-ary RR-LOCO code  along both wordline and bitline directions are very similar, and the designed codes have the same overall rate (including bridging symbols). The 1D $4$-ary RR-LOCO code  has a shorter overall block length corresponding to $12$ bits per coded page ($10$ symbols plus $2$ bridging symbols) in comparison to the 1D binary RR-LOCO code which has overall block length of $36$ bits on the coded page.  Moreover, in the code design, the 1D $4$-ary RR-LOCO encoder uses an adder size of $18$ bits, while the 1D binary RR-LOCO requires an adder size of $24$ bits. 




An examination of level probabilities induced by 1D binary and 1D $4$-ary RR constraints provides some intuitive insight into the experimental results in Figs.~\ref{exp_RR_all_life} and~\ref{exp_RR_high_PE}. The probabilities of binary symbols $0$ and $1$ under the RLL $(d,k)=(0,1)$ constraint are approximately $0.2764$ and $0.7236$, respectively \cite{siegel_nasit2015}. Asymptotically, this leads to probabilities of individual symbols corresponding to levels in 
$\mathcal{V}_0 =\{4,5,6,7\}$ and 
$\mathcal{V}_1=\{0,1,2,3\}$ of about $0.0691$ and $0.1809$, respectively. From the FSTD of the $4$-ary constraint forbidding patterns in $\mathcal{R}_4$, shown in  Fig.~\ref{fig_3}, we find that the probabilities of individual symbols corresponding to levels in 
$\mathcal{W}_0=\{6,7\}$, $\mathcal{W}_1=\{4,5\}$, $\mathcal{W}_2=\{2,3\}$, and 
$\mathcal{W}_3=\{0,1\}$ are about $0.0787$, $0.1030$, $0.1591$, and $0.1591$, respectively. Bridging symbols change these probabilities slightly, further increasing the probabilities of symbols corresponding to levels in $\{0,1,2,3\}$ relative to symbols corresponding to levels in $\{4,5,6,7\}$. These probabilities contrast with those of uncoded random data, where each symbol/level has the same probability of $1/8=0.125$. 

The modified symbol probabilities help to explain the observed relative performances of the 1D~binary RR codes in the wordline and bitline directions along with the 2D RR code. Applying 1D binary RR coding in the wordline direction also indirectly reduces the probability of detrimental patterns in the bitline direction, and vice versa. This reduces the expected advantage of bitline coding over wordline coding in the presence of more severe ICI in the bitline direction. Similarly, the advantage of 2D coding over 1D coding in either direction is less expected (even without taking into account the rate penalty associated with 2D coding). 

We remark that the designed codes are efficient, with rates fairly close to capacity, and the symbol and pattern probabilities observed in the data written to the Flash memory are close to the theoretical values mentioned above.

The cross-over behavior observed in Fig.~\ref{exp_RR_all_life} can be explained if the level patterns eliminated by the code, especially ICI-prone patterns, are not the only significant contributors to error early in the device lifetime. The binary RR coding significantly changes level probabilities compared with the uncoded setting, possibly increasing the probability of some of the remaining level patterns that cause errors due to other effects, and accordingly increasing their contribution to the BER at low P/E cycles. One suggestion to prevent this behavior is to apply  different constraints before and after the cross-over point. The  reconfigurability feature of LOCO code designs could be exploited, and a machine learning module could be used to identify the device status and direct the transition from one code to another at the appropriate time based on that status. In this regard, we also note that machine learning modeling, as proposed in~\cite{simeng_codeaware}, can be used to characterize the spatio-temporal ICI effects of the Flash memory device and provide a tool for optimizing the design of RR-LOCO codes.

\section{Conclusion}\label{sec_concl}

We introduced read-and-run (RR) constrained coding schemes for modern Flash devices. RR coding schemes eliminate patterns prone to ICI-induced errors while allowing systematic encoder and decoder implementations, high overall rates, and page separation in data recovery. We analyzed properties of 1D binary RR-LOCO codes, 1D $4$-ary RR-LOCO codes, and a 2D binary RR code. The three RR coding schemes offer different advantages, and we suggest that system requirements at different stages of the device lifetime should determine the most suitable scheme or schemes to use. Experimental results reveal significant P/E-cycle lifetime gains in a commercial Flash device. Future work includes the incorporation of LDPC codes~\cite{ahh_md} with RR coding schemes and the  development of machine learning-aided,  reconfigurable RR coding schemes  to maximize Flash device lifetime.



\end{document}